\documentclass[journal]{IEEEtran}

\usepackage{graphicx}
\usepackage[toc]{glossaries}

\usepackage{float}
\restylefloat{table}

\usepackage[hidelinks]{hyperref}
\hypersetup{
    colorlinks=true,
    linkcolor=blue,
    filecolor=magenta,      
    urlcolor=cyan,
}

\usepackage{amsmath}
\usepackage{amssymb}
\usepackage{bbm}
\usepackage{eucal}
\usepackage{booktabs}
\usepackage{caption,subcaption}
\usepackage{changes}
\usepackage{pifont}
\usepackage{rotating}
\usepackage{framed}
\usepackage[numbers]{natbib}
\usepackage{makecell}
\usepackage{float}

\usepackage{placeins}

\usepackage{tikz}
\usetikzlibrary{backgrounds}

\let\labelindent\relax
\usepackage{enumitem}

\DeclareMathAlphabet\mathbfcal{OMS}{cmsy}{b}{n}

\usepackage{tcolorbox}
\usepackage{nicefrac,xfrac}

\DeclareMathOperator*{\softmax}{\textsc{SoftMax}}

\newcommand{\figref}[1]{Figure~\ref{#1}}
\newcommand{\tableref}[1]{Table~\ref{#1}}
\renewcommand{\eqref}[1]{Equation~\ref{#1}}
\newcommand{\secref}[1]{Section~\ref{#1}}

\newcommand{\mycomment}[1]{}

\usepackage{adjustbox}
\usepackage{array}
\usepackage{multirow}
\newcolumntype{L}[1]{>{\raggedright\let\newline\\\arraybackslash\hspace{0pt}}m{#1}}
\newcolumntype{C}[1]{>{\centering\let\newline\\\arraybackslash\hspace{0pt}}m{#1}}
\newcolumntype{R}[1]{>{\raggedleft\let\newline\\\arraybackslash\hspace{0pt}}m{#1}}

\newcommand{\cmark}{\text{\ding{51}}}%
\newcommand{\xmark}{\text{\ding{55}}}%

\newcolumntype{R}[2]{%
    >{\adjustbox{angle=#1,lap=\width-(#2)}\bgroup}%
    c%
    <{\egroup}%
}

\makeglossaries

\newacronym{drs}{DRS}{Diabetic Retinopathy Screening}
\newacronym{vqa}{VQA}{Visual Question Answering}
\newacronym{slam}{SLAM}{Simultaneous Localization and Mapping}
\newacronym[plural=CNNs,firstplural=Convolutional Neural Networks (CNNs)]{cnn}{CNN}{Convolutional Neural Network}
\newacronym{nlp}{NLP}{natural language processing}
\newacronym{mcb}{MCB}{Multimodal Compact Bilinear}
\newacronym{mlb}{MLB}{Multimodal Low-rank Bilinear}
\newacronym{mutan}{MUTAN}{Multimodal Tucker Fusion for Visual Question Answering}
\newacronym{idrid}{IDRID}{Indian Diabetic Retinopathy Image Dataset}
\newacronym[plural=RNNs,firstplural=Recurrent Neural Networks (RNNs)]{rnn}{RNN}{Recurrent Neural Network}
\newacronym{lstm}{LSTM}{Long Short-term Memory}
\newacronym{bow}{BOW}{bag-of-words}
\newacronym{gru}{GRU}{Gated Recurrent Units}
\newacronym[plural=QAs,firstplural=Questions \& Answers (QAs)]{qa}{QA}{Question \& Answer}
\newacronym{ma}{MA}{Microaneurysms}
\newacronym{he}{HE}{Hemorrhages}
\newacronym{ex}{EX}{Hard Exudates}
\newacronym{se}{SE}{Soft Exudates}
\newacronym{nar}{NAR}{no apparent retinopathy}
\newacronym{dr}{DR}{Diabetic Retinopathy}
\newacronym{dme}{DME}{Diabetic Macular Edema}
\newacronym{pdr}{PDR}{Proliferative Diabetic Retinopathy}
\newacronym{wmlb}{WMLB}{Weighted Multimodal Low-rank Bilinear Attention Network}
\newacronym{dl}{DL}{deep learning}
\newacronym{qcmlb}{QC-MLB}{Question-Centric Multimodal Low-rank Bilinear}
\newacronym{bert}{BERT}{Bidirectional Encoder Representations from Transformers}
\newacronym{bleu}{BLEU}{Bilingual Evaluation Understudy}
\newacronym{mlm}{MLM}{Masked Language Model}
\newacronym{nsp}{NSP}{Next Sentence Prediction}
\newacronym{relu}{\textsc{ReLU}}{rectified linear unit}
\newacronym{nn}{NN}{neural network}
\newacronym{chal}{ImageCLEF-VQA-Med}{ImageCLEF-VQA-Med}
\newacronym{proposed}{\textit{\textless~Model~\textgreater}}{\textbf{Full name of the proposed model}}
\newacronym{mri}{MRI}{magnetic resonance imaging}
\newacronym{brats}{BraTS19}{Brain Tumors in Multimodal Magnetic Resonance Imaging Challenge 2019}
\newacronym{kits}{KiTS19}{Kidney Tumor Segmentation Challenge 2019}
\newacronym{ibsr}{IBSR18}{Internet Brain Segmentation Repository}
\newacronym{hene}{U-HAND}{Ume{\aa} Head and Neck Database}
\newacronym{pros}{U-PRO}{Ume{\aa} Pelvic Region Organs}
\newacronym{hpc2n}{HPC2N}{High Performance Computer Center North}
\newacronym{flops}{FLOPs}{floating point operations}

\newacronym{ct}{CT}{computed tomography}
\newacronym{t1c}{T1c}{post-contrast T1-weighted}
\newacronym{t2}{T2w}{T2-weighted}
\newacronym{t1}{T1w}{T1-weighted}
\newacronym{flair}{FLAIR}{T2  Fluid  Attenuated  Inversion  Recovery}
\newacronym{lgg}{LGG}{low grade glioma}
\newacronym{hgg}{HGG}{high grade glioma}

\newacronym[plural=DSCs]{dsc}{DSC}{Dice similarity coefficient}
\newacronym{seb}{SEB}{Squeeze-and-Excitation block}
\newacronym[plural=REBs,firstplural=ResNet blocks]{res}{REB}{ResNet block}
\newacronym{hd95}{HD95}{$95^{th}$ percentile of the Hausdorff distance}
\newacronym{dauc}{DAUC}{Dice Area Under Curve}
\newacronym{rftp}{RFTPs}{Ratio of Filtered True Positives}

\newacronym[plural=SDs,firstplural=standard deviations (SDs)]{sd}{SD}{standard deviation}

\newacronym{ce}{CE}{categorical cross--entropy}
\newacronym{gpu}{GPU}{Graphical Processing Unit}

\usepackage{scalefnt}

\definecolor{wrongColor}{RGB}{229,57,53}
\definecolor{rightColor}{RGB}{56,142,60}

\begin{document}
%


\title{Evaluation of Multi-Slice Inputs to Convolutional Neural Networks for Medical Image Segmentation}

\author{Minh H. Vu$^{1}$,
        Guus Grimbergen$^{2}$,
        Tufve Nyholm$^{1}$,
        and Tommy L\"{o}fstedt$^{1}$
\thanks{$^{1}$M. Vu, T. Nyholm, and T. L\"{o}fstedt are with the Department of Radiation Sciences, Ume{\aa} University, Ume{\aa}, Sweden. E-mail: minh.vu@umu.se.}
\thanks{$^{2}$G. Grimbergen is with the Department of Biomedical Engineering, Eindhoven University of Technology, 5612 AZ Eindhoven, the Netherlands.}
}
\maketitle              

\begin{abstract}
When using Convolutional Neural Networks (CNNs) for segmentation of organs and lesions in medical images, the conventional approach is to work with inputs and outputs either as single slice (2D) or whole volumes (3D). One common alternative, in this study denoted as pseudo-3D, is to use a stack of adjacent slices as input and produce a prediction for at least the central slice. This approach gives the network the possibility to capture 3D spatial information, with only a minor additional computational cost.

In this study, we systematically evaluate the segmentation performance and computational costs of this pseudo-3D approach as a function of the number of input slices, and compare the results to conventional end-to-end 2D and 3D CNNs. The standard pseudo-3D method regards the neighboring slices as multiple input image channels. We additionally evaluate a simple approach where the input stack is a volumetric input that is repeatably convolved in 3D to obtain a 2D feature map. This 2D map is in turn fed into a standard 2D network. We conducted experiments using two different CNN backbone architectures and on five diverse data sets covering different anatomical regions, imaging modalities, and segmentation tasks.

We found that while both pseudo-3D methods can process a large number of slices at once and still be computationally much more efficient than fully 3D CNNs, a significant improvement over a regular 2D CNN was only observed for one of the five data sets.
An analysis of the structural properties of the segmentation masks revealed no relations to the segmentation performance with respect to the number of input slices.

The conclusion is therefore that in the general case, multi-slice inputs appear to not significantly improve segmentation results over using 2D or 3D CNNs.
\end{abstract}


\section{Introduction}
\label{sec:intro}

Segmentation of organs and pathologies are common activities for radiologists and routine work for radiation oncologists.
Nowadays manual annotation of such regions of interest is aided by various software toolkits for image enhancement, automated contouring, and structure analysis in all fields on image-guided radiotherapy~\citep{rangayyan2007review, bauer2013survey, dolz2016interactive}. Over the recent years, \gls{dl} has emerged as a very powerful concept in the field of medical image analysis. The ability to train complex neural networks by example to independently perform a vast spectrum of annotation tasks has proven itself a promising method to produce segmentations of organs and lesions with expert-level accuracy~\cite{shen2017deep, litjens2017survey}.

For both organ segmentation and lesion segmentation, the most common \gls{dl} model is the \gls{cnn}. Whereas the classic approach of segmenting 3D medical volumes by \glspl{cnn} consists of training on and predicting the individual 2D slices independently, the interest has shifted in recent years towards full 3D convolutions in vo1umetric neural networks~\citep{sahiner2019deep, litjens2017survey, milletari2016v, cciccek20163d, dou20173d}. Volumetric convolution kernels have the advantage of taking inter-slice context into account, thus preserving more of the spatial information than what is possible when using 2D convolutions within slices. However, volumetric operations require a much larger amount of computational resources. For medical image applications, the lack of sufficient \gls{gpu} memory to fit entire volumes at once requires in almost all cases a patch-based approach, reduced input sizes, and/or small batch sizes and therefore longer training times.

\subsection{Related Work}
\label{subsec:related}

In terms of fully connected, end-to-end 3D networks, studies often attempt to compensate for the small patch size that can maximally fit into the \gls{gpu} memory at once by creating more efficient architectures or utilizing post-processing methods. The original U-Net by \citet{ronneberger2015u}, an architecture which was, at that time, and still is, a popular and powerful network for semantic medical image segmentation, was first reintroduced as a 3D variant by \citet{cciccek20163d}. The 3D U-Net was used by \citet{minh2019end,minh2019tunet} in a cascaded approach where a first coarse prediction was used to generate a candidate region in which a second, finer-grained prediction was performed; this proved to be an effective way of reducing the amount of input data for the final prediction. V-Net by \citet{milletari2016v} extended the network of \cite{cciccek20163d} by adding residual connections to the 3D U-Net.

\citet{li2017compactness} reduced the computational cost required for a fully connected 3D \gls{cnn} by replacing the deconvolution steps in the upsampling phase with dilated convolution to preserve the spatial resolution of the feature maps. VoxResNet~\citep{chen2018voxresnet} is a very deep residual network that was trained on small 3D patches. The resulting output probability map was combined with the original multimodal volumes into a second VoxResNet to obtain a more accurate output. A related approach from \citet{yu2017volumetric} extended this architecture by implementing long residual connections between residual blocks, in addition to the short connections within the residual blocks. The same group proposed another densely connected architecture called DenseVoxNet~\citep{yu2017automatic}, where each layer had access to the feature maps of all its preceding layers, decreasing the number of parameters and possibly avoiding to learn redundant feature maps.

\citet{Lu2017} used a graph cut model to refine the output of their coarse 3D \gls{cnn}. A 3D network composed of two separate convolutional pathways, at low and high resolution, was introduced by \citet{kamnitsas2017efficient}. For improvement, the resulting segmentation was, in turn, post-processed by a Conditional Random Field. A variant of this multi-scale feature extraction during convolution was used by \citet{lian2018multi}, who used this procedure in the encoding phase of their U-Net-like 3D \gls{cnn}. \citet{ren2018interleaved} exploited the small size of regions of interest in the head and neck area (\textit{i.e.} the optic nerves and chiasm) to build an interleaved combination of small-input, shallow \glspl{cnn} trained at different scales and in different regions. \citet{feng2019deep} used a two-step procedure: a first 3D U-Net was used to localize thoracic organs in a substantially downsampled volume, and crop to a bounding box around each organ. Then, individual 3D U-Nets were trained to segment each organ inside its subvolume at the original resolution. Another example of 3D convolutions applied only on a small region of interest is from the work of \citet{anirudh2016lung}, who randomly sampled subvolumes in lung images for which the centroid pixel intensity was above a certain intensity threshold, to classify the subvolume as containing a lung nodule or not.

While these studies have shown that 3D \glspl{cnn} are worth the effort, alternative approaches have been investigated to involve volumetric context to improve segmentation while avoiding 3D convolutions altogether. One of the more common methods, usually called 2.5D, is to use \glspl{cnn} that combine tri-planar 2D \glspl{cnn} from intersecting orthogonal patches \citep{prasoon2013deep, roth2014new, de2015deep, yang2015automated, lyksborg2015ensemble, mlynarski20193d, kitrungrotsakul2019vesselnet, geng20192}. This can be a computationally efficient way of incorporating more 3D spatial information, and these studies all present promising results. However, this method is limited in the volumetric information it can encompass at once.

We, therefore, investigate a method that uses a volumetric input but is still largely 2D based with a minimal amount of 3D operations. Instead of a method that takes a single 2D slice as input, and outputs the 2D segmentation of that slice, one can also incorporate neighboring slices to provide a 3D context to enhance segmentation performance. A common approach to this is to include neighboring slices to a central slice as multiple input image channels. \citet{novikov2018deep} included the preceding and succeeding axial slice for vertebrae and liver segmentation. Such a three-slice input was also used by \citet{kitrungrotsakul2019cascade} for the detection of mitotic cells in 4D data (spatial + temporal). This was a cascaded approach where a first detection step with a three-slice input produced results for these three slices. In the second step, they reduced the number of false positives where for each slice the time-frame before and after was included. In a deep \gls{cnn} for liver segmentation, \citet{han2017automatic} used five neighboring slices. \citet{ghavami2018integration} compared incorporating three, five, and seven slices for prostate segmentation from ultrasound images. While their method produced promising segmentation results, no significant difference was found between these three input sizes. In a recent paper, \citet{ganaye2019removing} employed a seven-slice input producing an output for the three central slices, which the authors refer to as 2.5D. This model was used to evaluate a loss function that penalized anatomically unrealistic transitions between adjacent slices. The authors did not report a significant improvement between the baseline 2D and 2.5D models, but the 2.5D model did outperform in terms of Hausdorff Distance when the non-adjacency loss was employed.

\subsection{Contributions}
\label{subsec:contrib}

In this paper, we systematically investigate using multiple adjacent slices as input to predict for the central slice in that subset, and we investigate this on the segmentation task in medical images. We will henceforth refer to any method based on this principle as \textit{pseudo-3D}. We compare the segmentation performance of a range of input multi-slice sizes ($d\in\{3, 5, \ldots, 13\}$) to conventional end-to-end 2D and fully 3D input-output \glspl{cnn}. We employ the common approach from the literature where each neighboring slice is put as a separate channel in the input, and we will refer to this method as \textit{the channel-based method}. Further, we introduce a second pseudo-3D method, that appears to have not been proposed in the literature before. This pseudo-3D method consists of two main components: a transition block that transforms a $d$-slice input into a single-slice (\textit{i.e.}~2D) feature map by using 3D convolutions, and this feature map is then followed by a standard 2D convolutional network, such as the U-Net~\citep{ronneberger2015u} or the SegNet~\citep{badrinarayanan2017segnet}, that produces the final segmentation labels. This method shall be referred to as \textit{the proposed method}. 

The main contributions of our work are:
\begin{enumerate}[noitemsep,nolistsep]
    \item We systematically compare the segmentation performance of 2D, pseudo-3D (with varying input size, $d$), and 3D approaches.

    \item We introduce a novel pseudo-3D method, using a transition block that transforms a multi-slice subvolume into a 2D feature map that can be processed by a 2D network. This method is compared to the channel-based pseudo-3D method.

    \item We compare the computational efficiency of fully 2D and 3D \glspl{cnn} to the pseudo-3D methods in terms of graphical memory use, number of model parameters, \gls{flops}, training time, and prediction time.

    \item We conduct all experiments on a diverse range of data sets, covering a broad range of data set sizes, imaging modalities, segmentation tasks, and body regions.
\end{enumerate}

\section{Proposed Method}
\label{sec:approach}

\begin{figure}[!t]
    \centering
    \includegraphics[width=.48\textwidth]{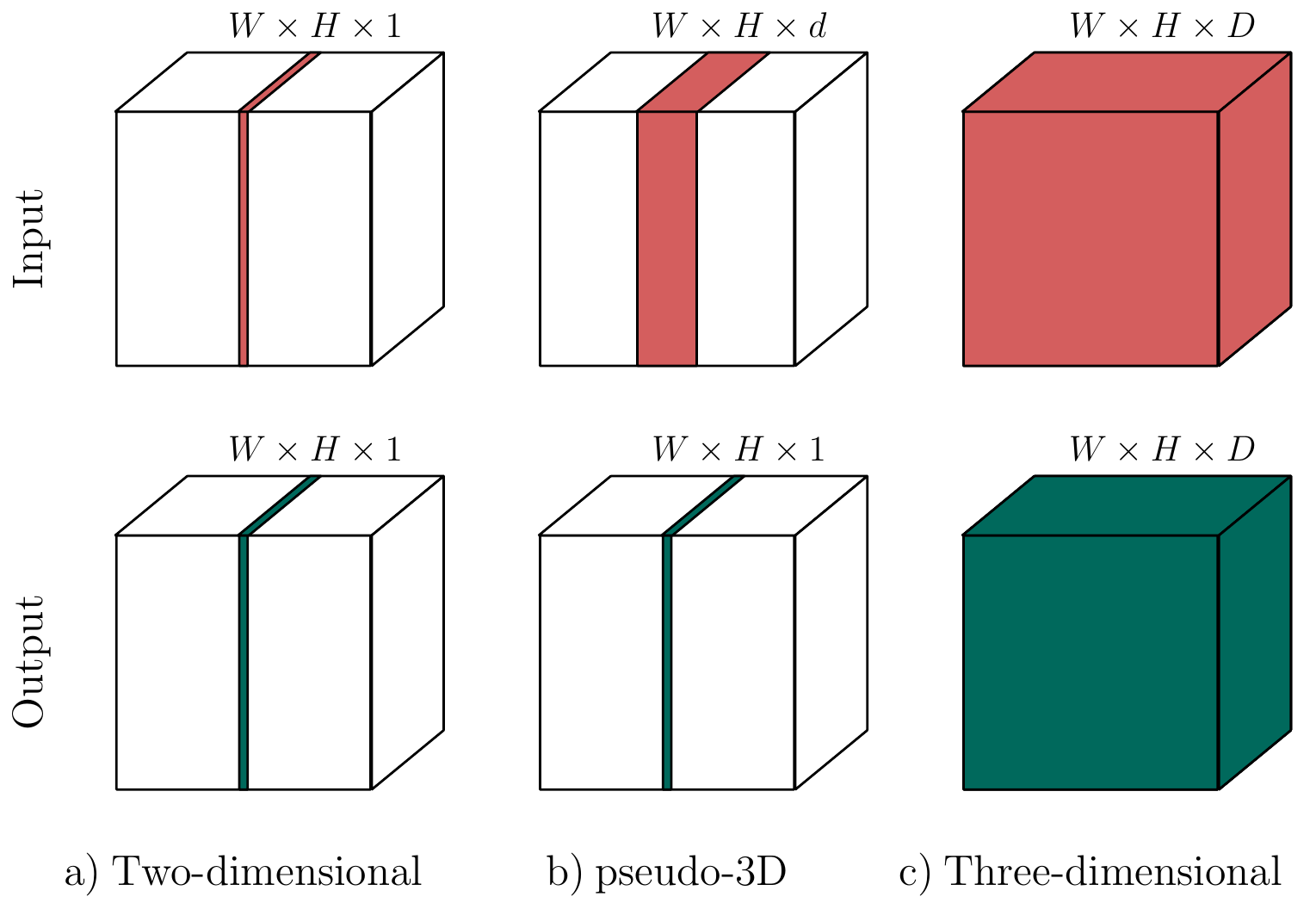}
    \caption{A comparison of 2D, pseudo-3D and 3D approaches. With a 2D network, the volume is segmented with a single slice input and output. Pseudo-3D uses multiple adjacent slices as input to produce an output of the central slice from the input. 3D approaches take in the whole volume at once, and return a prediction for the whole volume as well. In the figure, the $W$, $H$, and $D$ are the original width, height, and depth of the input volume, respectively.
    }
    \label{fig:compared}
\end{figure}

The underlying concept of the pseudo-3D methods is similar to that of standard slice-by-slice predictions using 2D \glspl{cnn}, but the input is now a subvolume with an odd number of slices, $d$, extracted from the whole volume with a total of $D$ slices. The output of the model is compared to the ground truth of the central slice. If $d=1$, the method is equivalent to a 2D \gls{cnn}. A fully 3D \gls{cnn} would be $d=D$ for both input and output, where all operations in the network are in 3D and the output volume is compared to the ground truth of the whole volume. See \figref{fig:compared} for an illustration of the proposed method. In this study, the number of slices in the input subvolume ranged from $d=3$ to $d=13$. In order to isolate the contribution of using multi-slice inputs, this work did not include multi-slice outputs---where the multiple outputs for each slice are usually aggregated using \textit{e.g.}~means or medians.

Let the input volume be of width $W$, height $H$, depth $d$, and have $C$ \textit{channels}. A common way of utilizing depth information to train with regards to the central slice is as follows: group the channel and depth dimension together as one, and consider the input to be of shape $W \times H \times (d\cdot C)$, \textit{i.e.}~with $d\cdot C$ channels. By incorporating the slices in the channel dimension, the multi-slice input can be processed by a regular 2D network. As was mentioned in \secref{subsec:contrib}, this method is denoted here as the \textit{channel-based method}.

The channel-based method is compared to a novel pseudo-3D approach denoted \textit{the proposed method}. Consider the input to be of shape $W \times H \times d \times C$. This is fed through a transition block with $L=\lfloor \frac{d}{2} \rfloor$ layers (where $\lfloor \cdot \rfloor$ is the floor function). In each layer, a 3D convolution with a kernel of size $3 \times 3 \times 3$ is applied to the volume within the image, after it has been padded in the $W$ and $H$ dimensions, but not in the $d$ dimension. Thus, after each layer in the transition block, the depth of the image is reduced by $2$ slices, while the width and height stay the same size. After the final convolution, the depth dimension is removed. Hence, the shapes change as
\begin{align*}
  W \times H \times d \times C &\rightarrow W \times H \times (d-2) \times C \\
                               &\rightarrow W \times H \times (d-4) \times C \\
                               &\rightarrow \cdots \\ 
                               &\rightarrow W \times H \times 3 \times C \\
                               &\rightarrow W \times H \times 1 \times C \\
                               &\rightarrow W \times H \times C.
\end{align*}

In both the proposed method and the channel-based method, the output layer of the network is the segmentation mask, with an output shape of $W \times H \times 1$. Hence, it produces a single segmentation slice, corresponding to the central slice of the input subvolume. See \figref{fig:compared} for an illustration of this.

\begin{figure*}[!th]
    \centering
    \begin{subfigure}[t]{0.7\textwidth}    
    \includegraphics[width=\textwidth]{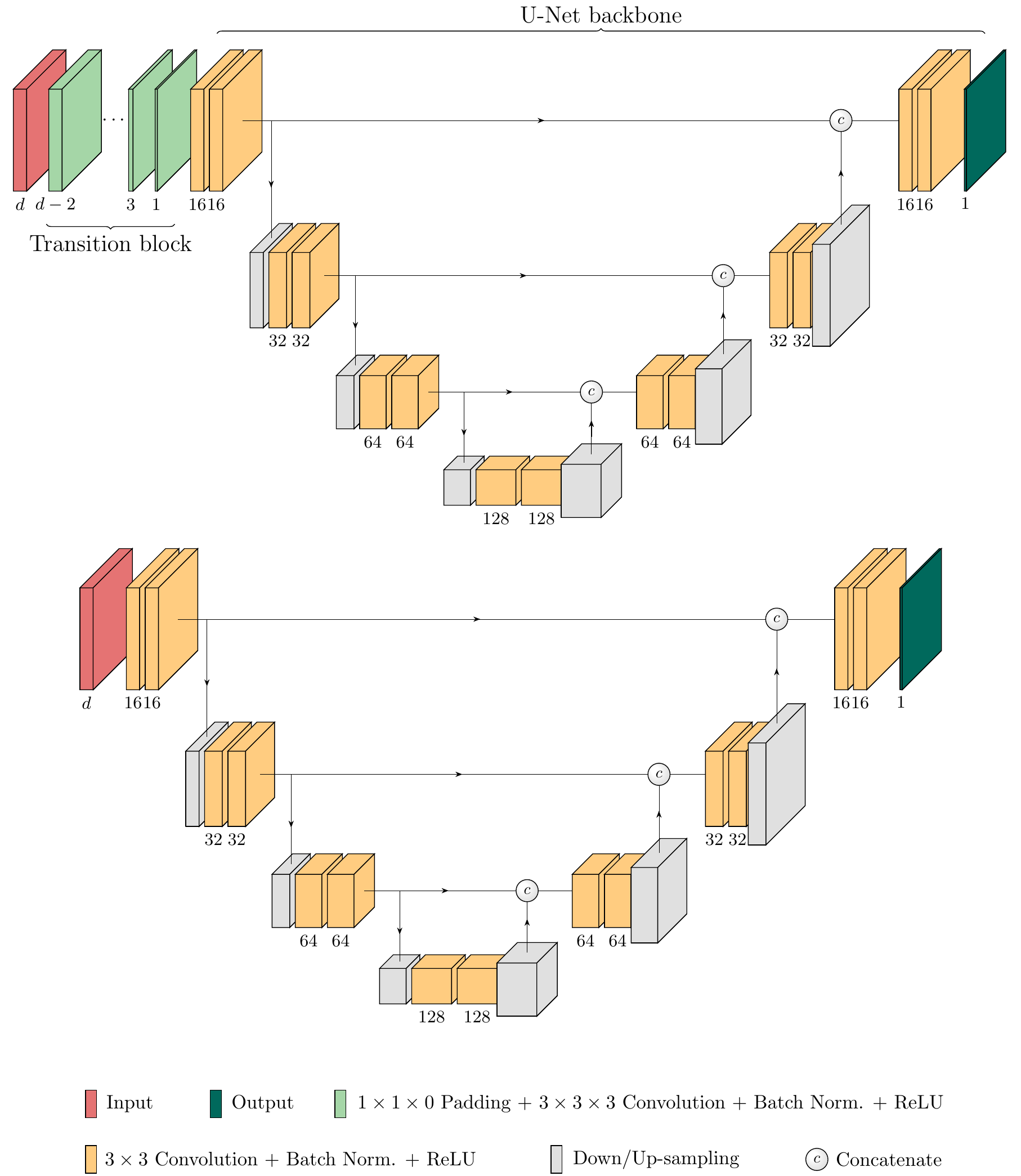}
    \end{subfigure}
    \caption{The proposed methods illustrated with the U-Net backbone. The output is the prediction for the central slice of the input. The numbers in the transition block indicate the depth and in the backbone the number of filters. Top: The Proposed Method where the transition block uses 3D convolutions and 2D padding to iteratively reduce the input from depth $d$ to $1$, while the width and height remain. Bottom: The Channel-Based method, where neighboring slices are input as separate channels, and the input can be fed into a 2D \gls{cnn} right away.}
    \label{fig:proposed}
\end{figure*}

The network architectures that were evaluated in this work was the U-Net~\citep{ronneberger2015u} and the SegNet~\citep{badrinarayanan2017segnet}, two popular variants of encoder-decoder architectures that have been successful in semantic medical image segmentation. An illustration of both pseudo-3D methods, with U-Net as the main network architecture, is given in \figref{fig:proposed}. Another illustration of the networks with SegNet backbone can be seen in \figref{fig:proposed_seg} in the Supplementary Material.

We evaluate the two pseudo-3D methods for $d \in \{3, 5, 7, 9, 11, 13\}$ and compare them to the corresponding conventional end-to-end 2D and 3D networks, all with the U-Net or SegNet architectures. This yields a total of $14$ different experiments for each data set (six input sizes for the two pseudo-3D methods, plus 2D and 3D methods, all with two network architectures). Apart from the segmentation performance, the computational cost is also evaluated across experiments in terms of the number of network parameters, the maximum required amount of \gls{gpu} memory, the number of \gls{flops}, the training time per epoch, and the prediction time per sample.


\section{Experiments}
\label{sec:exp}

We here present the data sets the experiments were conducted on, as well as the encompassing information and parameters used in the experiments.

\subsection{Materials}
\label{subsec:material}

To test the generalizing capabilities of the methods, we ran experiments on five different data sets, covering a variety of modalities, data set sizes, segmentation tasks, and body areas. Three of the data sets are publicly available, as they were part of segmentation challenges. On top of those, we further used two in-house data sets collected at the University Hospital of Ume\aa{}, Ume\aa{}, Sweden.

\subsubsection{\acrlong{pros}}
\label{subsubsec:prostate}

\begin{figure}[!th]
    \centering
    \includegraphics[trim=5 0 5 0, clip, width=.48\textwidth, height=.24\textwidth]{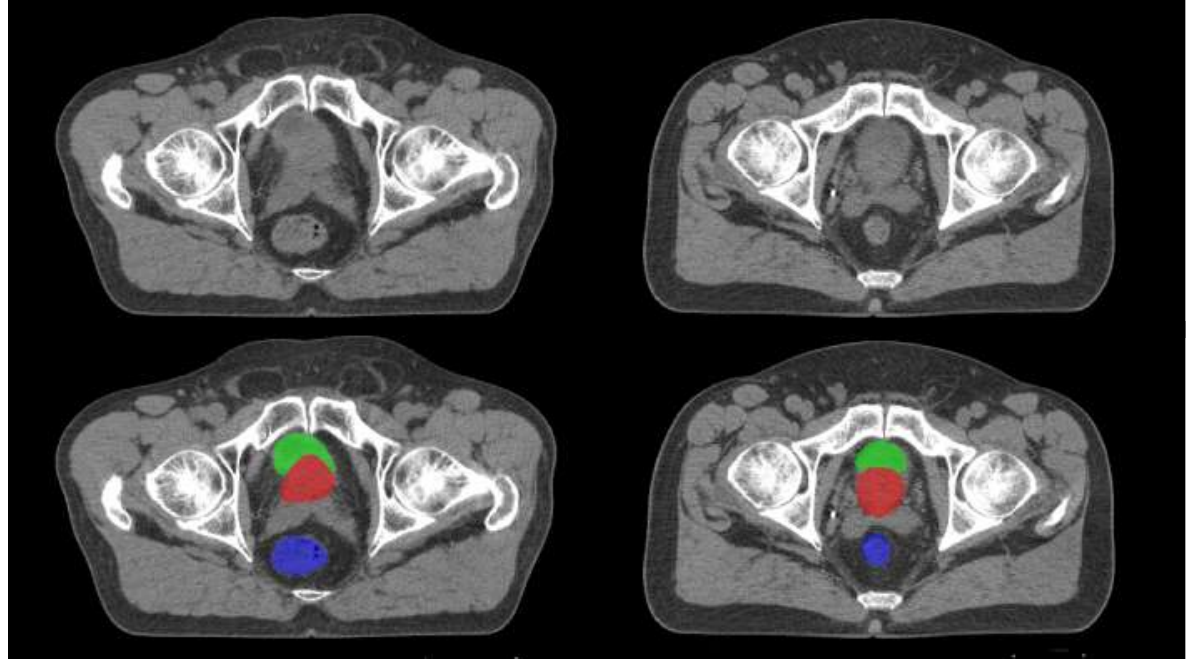}
    \caption{\acrlong{pros} data set. From top to bottom: images and ground truth images of the prostate (red), bladder (green) and rectum (blue).}
    \label{fig:pros}
\end{figure}

An in-house data set containing \gls{ct} images of the pelvis region from $1\,244$ patients that underwent radiotherapy for prostate cancer at the University Hospital of Ume{\aa}, Ume\aa, Sweden. We denote this data set \gls{pros}. The delineated structures include the prostate (in most cases annotated as the clinical or gross target volume) and some organs at risk, among them the bladder and rectum. The individual structure masks were merged into a single multilabel truth image, with pixel value $1$ for the prostate, $2$ for the bladder, and $3$ for the rectum (see \figref{fig:pros}). Patients without the complete set of structures were excluded, resulting in a final data set containing $1\,148$ patients.

\subsubsection{\acrlong{hene}}
\label{subsubsec:headneck}

\begin{figure}[!th]
    \centering
    \includegraphics[trim=15 0 15 0, clip, width=.48\textwidth, height=.24\textwidth]{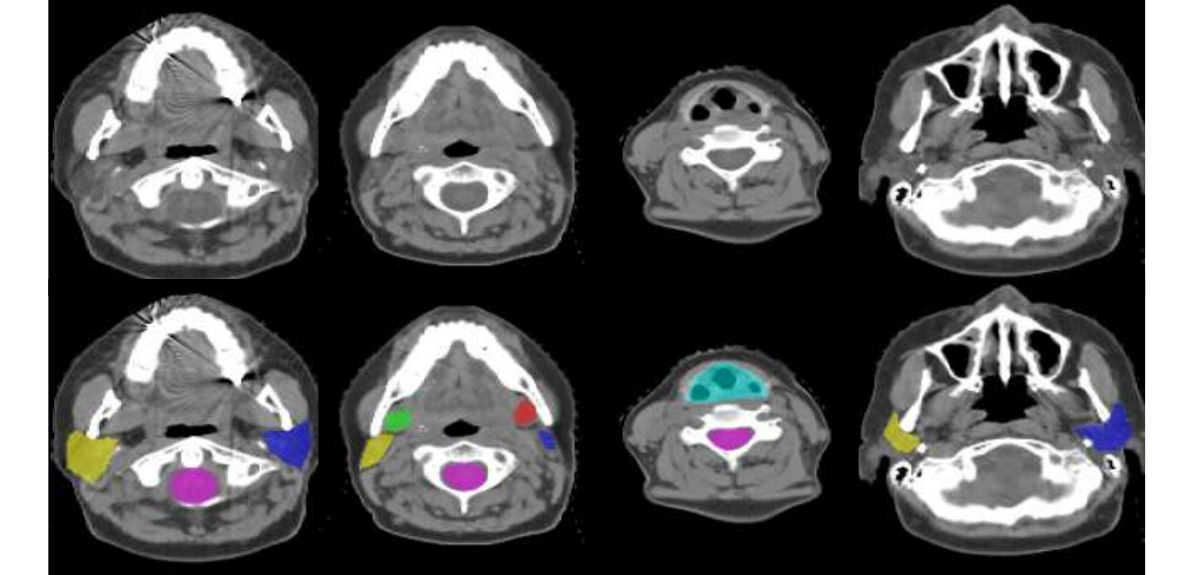}
    \caption{\acrlong{hene}. From top to bottom: images and ground truth images at different slices of the left and right submandibular glands (red and green), left and right parotid glands (dark blue and yellow), larynx (light blue), and medulla oblongata (pink).}
    \label{fig:hene}
\end{figure}

An in-house data set containing \gls{ct} images of the head and neck region of $110$ patients. This data set comprises the patients from the University Hospital of Ume{\aa}, Ume\aa{}, Sweden, that participated in the ARTSCAN study \citep{zackrisson2011two}. We denote this data set \gls{hene}. For each \gls{ct} image, manual annotations of the target volumes and various organs at risk were provided. The organ structures that were included with this data were the bilateral submandibular glands, bilateral parotid glands, larynx, and medulla oblongata (see \figref{fig:hene}). After removal of faulty \gls{ct} volumes where the slice spacing changed within a volume and excluding patients in which not all of the six aforementioned structures were present, the final data set contained $73$ patients.

\subsubsection{\acrlong{brats}}
\label{subsubsec:brats}

The \gls{brats} \citep{menze2014multimodal,bakas2017advancing} was part of the MICCAI 2019 conference. It contains multimodal pre-operative \gls{mri} data of $285$ patients with pathologically confirmed \gls{hgg} ($n=210$) or \gls{lgg} ($n=75$) from 19 different institutes. For each patient, \gls{t1}, \gls{t1c}, \gls{t2}, and \gls{flair} scans were available, acquired with different protocols and various scanners at $3$~T. 

\begin{figure}[!th]
    \centering
    \includegraphics[trim=15 0 15 0, clip, width=.48\textwidth, height=.24\textwidth]{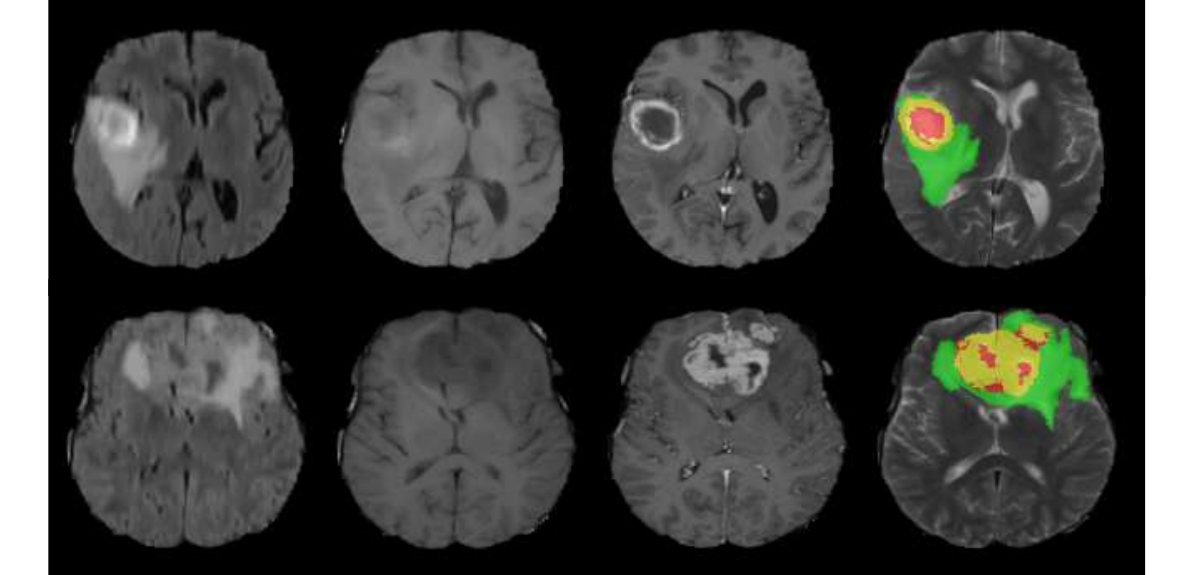}
    \caption{Manual expert annotation of two patients with \gls{hgg} from the \acrlong{brats} data set. Shown are image patches with the tumor structures that are annotated in the different modalities. The image patches show (from left to right): (1) the whole tumor visible in \gls{flair}, (2-3) the enhancing and tumor structures visible in \gls{t1} and \gls{t1c}, respectively, and (4) the final labels visible in \gls{t2}. The segmentations are combined to generate the final labels of the tumor structures:  the necrotic and non-enhancing tumor core (NCR/NET---label 1, red), the peritumoral edema (ED---label 2, green) and the GD-enhancing tumor (ET---label 4, yellow).} 
    \label{fig:brats18}
\end{figure}

Manual segmentations were carried out by one to four raters and approved by neuroradiologists. The necrotic and non-enhancing tumor core, peritumoral edema, and contrast-enhancing tumor were assigned labels $1$, $2$, and $4$ respectively (see \figref{fig:brats18}). The images were co-registered to the same anatomical template, interpolated to a uniform voxel size and skull-stripped.

\subsubsection{\acrlong{kits}}
\label{subsubsec:kits}

\begin{figure}[!th]
    \centering
    \includegraphics[trim=15 0 15 0, clip, width=.48\textwidth, height=.24\textwidth]{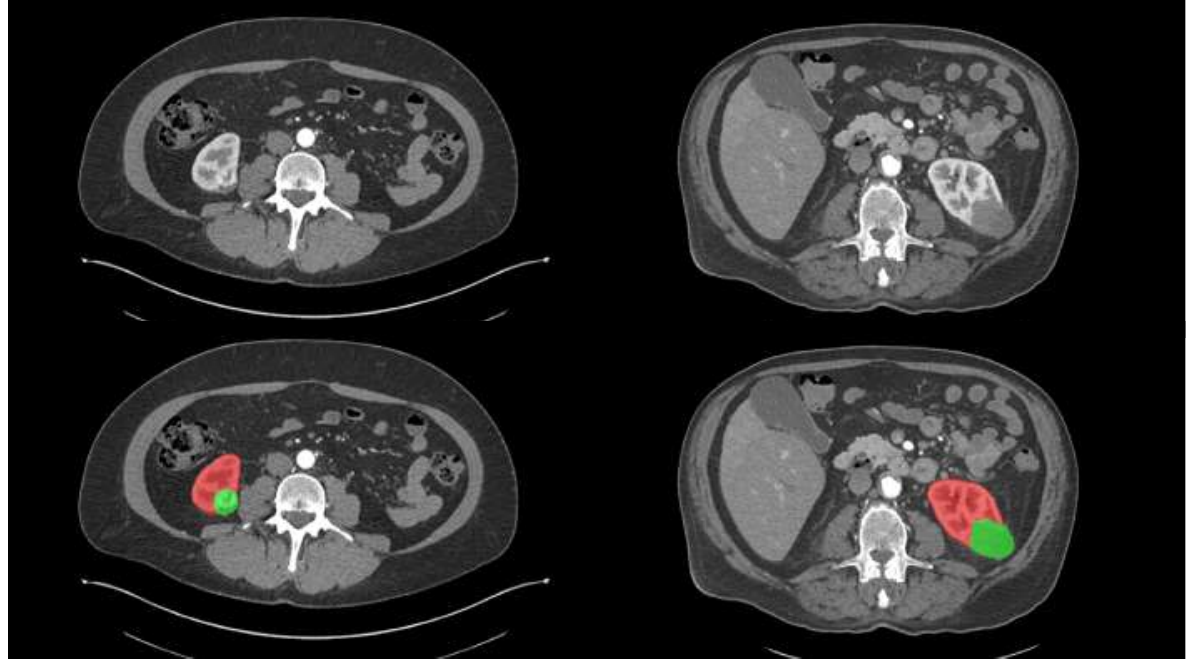}
    \caption{\acrlong{kits} data set. From top to bottom: images and ground truth images of the kidney (red) and kidney tumor (green).}   
    \label{fig:kits19}
\end{figure}

The data set for the \gls{kits} challenge \citep{heller2019kits19}, part of the MICCAI 2019 conference, contains preoperative \gls{ct} data from $210$ randomly selected kidney cancer patients that underwent radical nephrectomy at the University of Minnesota Medical Center between 2010 and 2018. Medical students annotated under supervision the contours of the whole kidney including any tumors and cysts (label 1), and contours of only the tumor component excluding all kidney tissue (label 2) (see \figref{fig:kits19}). Afterward, voxels with a radiodensity of less than $-30$~HU were excluded from the kidney contours, as they were most likely perinephric fat. 

\subsubsection{\acrlong{ibsr}}
\label{subsubsec:ibsr}

\begin{figure}[!ht] 
    \centering
    \includegraphics[trim=15 0 15 0, clip, width=.48\textwidth, height=.24\textwidth]{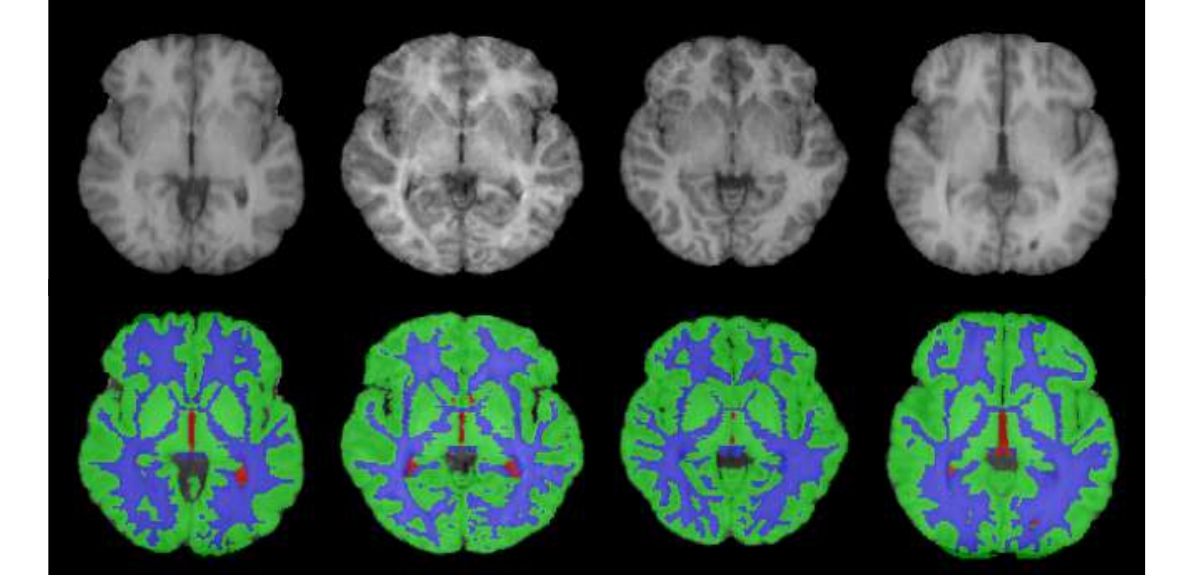}
    \caption{\acrlong{ibsr} data set. Axial slices of three patients with the ground truth of the cerebrospinal fluid (red), white matter (green) and gray matter (blue).}
    \label{fig:ibsr}
\end{figure}

The \gls{ibsr} data set \citep{cocosco1997brainweb} is a publicly available data set with $18$ \gls{t1} \gls{mri} volumes, and is commonly used as a standard data set for tissue quantification and segmentation evaluation. Whole-brain segmentations of cerebrospinal fluid (CSF), gray matter, and white matter were included with their respective labels $1$, $2$, and $3$ (see \figref{fig:ibsr}). 

\subsection{Preprocessing}
\label{subsec:prep}
Due to the diverse range of data sets, it must be ensured that the training data is as similar as possible across experiments in order to achieve a fair comparison.

\begin{table*}[!t]
\def\widthdetail{4.6 cm}
\def\width{2 cm}
\caption{Data sets and augmentation techniques in this study. $W$, $H$, and $D$ each denote that the volume shape is varied in width, height, and/or depth, respectively. $\alpha$, $\beta$, and $\gamma$ each denote that the voxel spacing is varied in width, height, and/or depth, respectively.}
\centering
\begin{tabular}{llrrrrr}
\toprule
material/data set                &  & \gls{brats}        & \gls{kits}                    & \gls{ibsr}        & \gls{hene}                    & \gls{pros}  \\ \midrule
type                            &  & \acrshort{mri}     & \acrshort{ct}                 & \acrshort{mri}    & \acrshort{ct}                 & \acrshort{ct}    \\ 
\#modalities                    &  & 4                  & 1                             & 1                 & 1                             & 1         \\
\#classes                       &  & 3                  & 2                             & 3                 & 6                             & 3         \\
\midrule
\#patients                      &  & 335                & 210                           & 18                & 73                            & 1\,148    \\
train                           &  & 268                & 168                           & 15-16             & 59                            &  734      \\
val                             &  &  67                &  42                           & 2-3               & 14                            &  184      \\
test                            &  &  67                &  42                           & 2-3               & 14                            &  230      \\
\midrule
original shape                  &  &  240-240-155       &  512-512-$D$                  &  256-128-256      & $W$-$H$-$D$                   &  512-512-$D$ \\
original voxel size (in mm)     &  &  1.0-1.0-1.0       &  $\alpha$-$\beta$-$\gamma$    &  1.0-1.0-1.0      & $\alpha$-$\beta$-$\gamma$     &  $\alpha$-$\beta$-$\gamma$ \\
preprocessed shape              &  &  160-192-128       &  256-256-128                  &  256-128-256      & 256-256-64                    &  256-256-128 \\
preprocessed voxel size (in mm) &  &  1.0-1.0-1.0       &  2.3-2.3-2.3                  &  1.0-1.0-1.0      & 1.3-1.0-5.8                   &  2.7-2.7-3.9 \\
\midrule
\multicolumn{7}{@{}l}{augmentation}  \\
\midrule
flip left-right         &  & \cmark         & \xmark        & \cmark        & \cmark        & \cmark        \\
elastic transform       &  & \cmark         & \cmark        & \cmark        & \cmark        & \cmark        \\
rotation                &  & \cmark         & \cmark        & \cmark        & \cmark        & \cmark        \\
shear                   &  & \cmark         & \cmark        & \cmark        & \cmark        & \cmark        \\
zoom                    &  & \cmark         & \cmark        & \cmark        & \cmark        & \cmark        \\
\bottomrule
\end{tabular}
\label{tab:datasetinfo}
\end{table*}

\subsubsection{Magnetic Resonance Image Preprocessing}
\label{subsubsec:mriprep}

The \gls{brats} and \gls{ibsr} data sets were N4ITK bias field corrected~\citep{tustison2010n4itk} and normalized to zero-mean and unit variance. The \gls{brats} volumes were cropped around the center to a resolution of $160 \times 192 \times 128$, to increase processing speed. This last step was skipped for the \gls{ibsr} data set because of the much smaller amount of data samples.

\subsubsection{Computed Tomography Image Preprocessing}
\label{subsubsec:ctprep}

\begin{figure}[!th] 
    \centering
    \includegraphics[width=.13\textwidth]{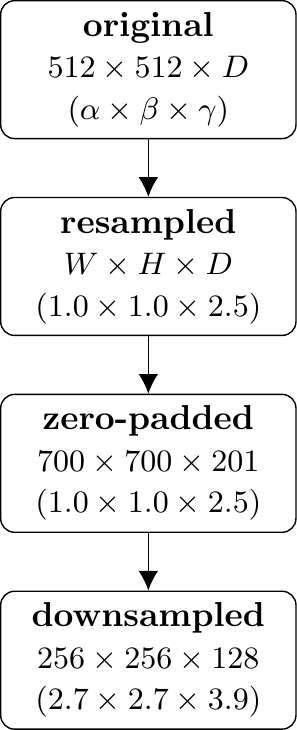}
    \caption{Preprocessing pipeline as applied on the \gls{pros} data set. Given are the resolutions and in parentheses the voxel dimensions in mm. $W$, $H$, and $D$ each denote that the volume shape is varied in width, height, and/or depth, respectively. $\alpha$, $\beta$, and $\gamma$ each denote that the voxel spacing is varied in width, height, and/or depth, respectively.}
    \label{fig:preprocess}
\end{figure}

In the \gls{pros}, \gls{hene}, and \gls{kits} data sets, all images had an $x \times y$ resolution (\textit{i.e.}~sagittal-coronal ) of $512 \times 512$ and a varying slice count. The voxel size also varied between patients, so a preprocessing pipeline (see \figref{fig:preprocess}) was set up to transform these two data sets to a uniform resolution and voxel size.

First, the data were resampled to an equal voxel size within the same set. The volumes were then zero-padded to the size of the single largest volume from that set after resampling. In order to increase processing speed and lower the memory consumption, the \gls{pros} and \gls{kits} volumes were thereafter downsampled to $256 \times 256 \times 128$, and the \gls{hene} volumes were downsampled to $256 \times 256 \times 64$. An example of this method pipeline is shown in \figref{fig:preprocess}.


As a final step, the images were normalized by clipping each case to the range $[-1000, 2000]$, subtracting $500$ and dividing by $1500$. 


\subsection{Training Details}
\label{subsec:training}

Our method was implemented in Keras 2.2.4\footnote{\url{https://keras.io}} using TensorFlow 1.12.0\footnote{\url{https://tensorflow.org}} as the backend. The experiments were trained on either a desktop computer with an NVIDIA RTX 2080~Ti \gls{gpu}, or the NVIDIA Tesla V100 GPUs from the \gls{hpc2n} at Ume{\aa} University, Sweden. Depending on the model, the convergence speed, and the data set size, a single experiment took from minutes to multiple days to complete.



\subsubsection{Experiment Setup}

For the 3D experiments, the \gls{brats} data set was the only data where the whole volumes could be fed into the network at once because of constraints in GPU memory. For the other data sets, we resorted to a patch-based approach where the input size would be $256 \times 256 \times 32$, the largest size possible for our available hardware.

In all experiments, we employed the Adam optimizer \citep{kingma2014adam} with an initial learning rate of $1 \cdot 10^{-4}$. If the validation loss did not improve after a certain number of epochs, we used a patience callback that dropped the learning rate by a factor of $0.2$ and an early stopping callback that terminated the experiment. Because of the differences in data set sizes, these callbacks had to be determined from initial exploratory experiments for each separate data set to ensure experiments did not run for too long or too short. The patience callbacks were set to five epochs for the \gls{brats}, \gls{kits}, and \gls{pros} experiments, six epochs for the \gls{hene} data set, and ten epochs for the \gls{ibsr} data set. The early stopping callbacks were set to $11$ epochs for \gls{pros} data, $12$ epochs for \gls{brats} and \gls{kits} data, $14$ for \gls{hene} data, and $25$ epochs for \gls{ibsr}. The maximum number of epochs an experiment could run for, regardless of any changes in the validation loss, was set to $100$ for the \gls{hene} and \gls{pros} data and $200$ for the other data sets. Batch normalization and a $L_2$ norm regularization, with parameter $1 \cdot 10^{-5}$, were applied to all convolutional layers, both in the transition block and in the main network. The \gls{relu} function was used as the intermediate activate function. The activation function of the final layer was the $\softmax$ function. Each data set was split into $80~\%$ training and $20~\%$ test set, and with the training set, in turn, being split into $80~\%$ for training and $20~\%$ for validation.

As loss function, we employed a combination of the \gls{dsc} and \gls{ce}. The \gls{dsc} is typically defined as
\begin{equation}\label{eq:dice}
    D(U, V)=\frac{2 \cdot |U \cap V|}{|U| + |V|}
\end{equation}
with $U$ the output segmentation and $V$ its ground truth. However, a differentiable version of \eqref{eq:dice}, the so-called soft \gls{dsc}, was used. The soft \gls{dsc} is defined as
\begin{equation}
    \mathcal{L}_{DSC}(u, v) = \frac{-2 \sum_i u_i v_i}{\sum_i u_i + \sum_i v_i + \epsilon},
\end{equation}
where for each label $i$, the $u$ is the $\softmax$ output of the network and $v$ is a one-hot encoding of the ground truth segmentation map. The $\epsilon$ is a small constant added to avoid division by zero.

The \gls{dsc} is a good objective for segmentation, as it directly represents the degree of overlap between structures. However, for unbalanced data sets with small structures and where the vast majority of pixels are background, it may converge to poorly generalizing local minima, since misclassifying only a few pixels can lead to large deviations in \gls{dsc}. A common way~\citep{roy2017relaynet, wong20183d} to resolve this is to combine the \gls{dsc} loss with the \gls{ce} loss, defined as
\begin{equation}
    \mathcal{L}_{CE}(u, v) = -\sum_i u_i \cdot \text{log} (v_i),
\end{equation}
and we did this as well. Hence, the final loss function was
\begin{equation} \label{eq:final_loss}
    \mathcal{L}(u, v) = \mathcal{L}_{dice}(u, v) + \mathcal{L}_{CE}(u, v).
\end{equation}

\subsubsection{Data Augmentation}

In order to artificially increase the data set size and to diversify the data, we employ various common methods for on-the-fly data augmentation: flipping along the horizontal axis, rotation within a range of $-1$ to $1$ degrees, shear images within the range of $-0.05$ to $0.05$, zoom with a factor between $0.9$ and $1.1$, and adding small elastic deformations as described in~\cite{simard2003best}. The data augmentation implementation we used was based on \cite{tensorlayer2017}.
The images in the \gls{kits} data are asymmetric along the $x$-axis because of the liver; therefore, vertical flipping was not applied on that data set as it would result in anatomically unrealistic images (see \tableref{tab:datasetinfo}).




\subsubsection{Evaluation}
\label{subsubsec:eval}

For evaluation of the segmentation performance, we employed the conventional \gls{dsc} as defined in \eqref{eq:dice}. In order to ensure a fair comparison and to investigate the variability of the results within experiments, we used five-fold cross-validation in each experiment (except for the \gls{pros}). Due to its much larger size, the experiments on the \gls{pros} data set were run only once.

To compare the computational cost of our proposed models to the corresponding 2D and 3D \gls{cnn} models, we extracted the number of trainable parameters, the maximum amount of \gls{gpu} memory used, the number of \gls{flops}, training time per epoch, and prediction time per sample.








\section{Results}
\label{sec:results}


\begin{table*}[!th]
\def\widthdetail{2.4 cm}
\def\width{1.9 cm}
\def\widthnarrow{1.6 cm}
\def\widthnarrowest{1 cm}
\caption{Architecture comparison. Experiments on U-Net architecture and multimodal \gls{brats} data set. Patch shape was set at $160 \times 192 \times d$ where $d$ is the number of slices. Here, \textit{t} and \textit{p} denote the training time per epoch and prediction time per sample, respectively.}
\centering
\begin{tabular}{llrrrrrr}
\toprule
Model                  &  & \#slices                  & \#params & memory                   & \acrshort{flops}           & \textit{t} per epoch  & \textit{p} per sample \\ 
\midrule
2D                     &  & 1                         & 493k   & 467MB                    & 2.450M            & 49s                   & 10.17s\\
\midrule
\multirow{6}{*}{proposed}
                       &  & 3                         & 495k   & 497MB                    & 2.463M            & 73s                   & 10.46s\\
                       &  & 5                         & 502k   & 519MB                    & 2.497M            & 88s                   & 11.11s\\
                       &  & 7                         & 509k   & 541MB                    & 2.532M            & 109s                  & 11.43s\\
                       &  & 9                         & 516k   & 563MB                    & 2.567M            & 156s                  & 12.15s\\
                       &  & 11                        & 523k   & 586MB                    & 2.602M            & 204s                  & 12.46s\\
                       &  & 13                        & 530k   & 601MB                    & 2.637M            & 241s                  & 14.73s\\
\midrule
\multirow{6}{*}{channel-based}
                       &  & 3                         & 493k   & 485MB                    & 2.451M            & 72s                   & 10.33s\\
                       &  & 5                         & 493k   & 497MB                    & 2.453M            & 82s                   & 10.49s\\
                       &  & 7                         & 493k   & 510MB                    & 2.454M            & 101s                  & 11.01s\\
                       &  & 9                         & 493k   & 523MB                    & 2.454M            & 138s                  & 11.17s\\
                       &  & 11                        & 494k   & 534MB                    & 2.457M            & 190s                  & 11.33s\\
                       &  & 13                        & 495k   & 541MB                    & 2.459M            & 249s                  & 12.39s\\
\midrule
3D                     &  & 128                       & 1\,461k   & 16\,335MB             & 7.306M            & 370s                  & 2.36s\\
\bottomrule
\end{tabular}
\label{tab:comparison}
\end{table*}

\begin{figure*}[!th]
    \centering
    \includegraphics[trim=110 0 100 40, clip, width=\textwidth]{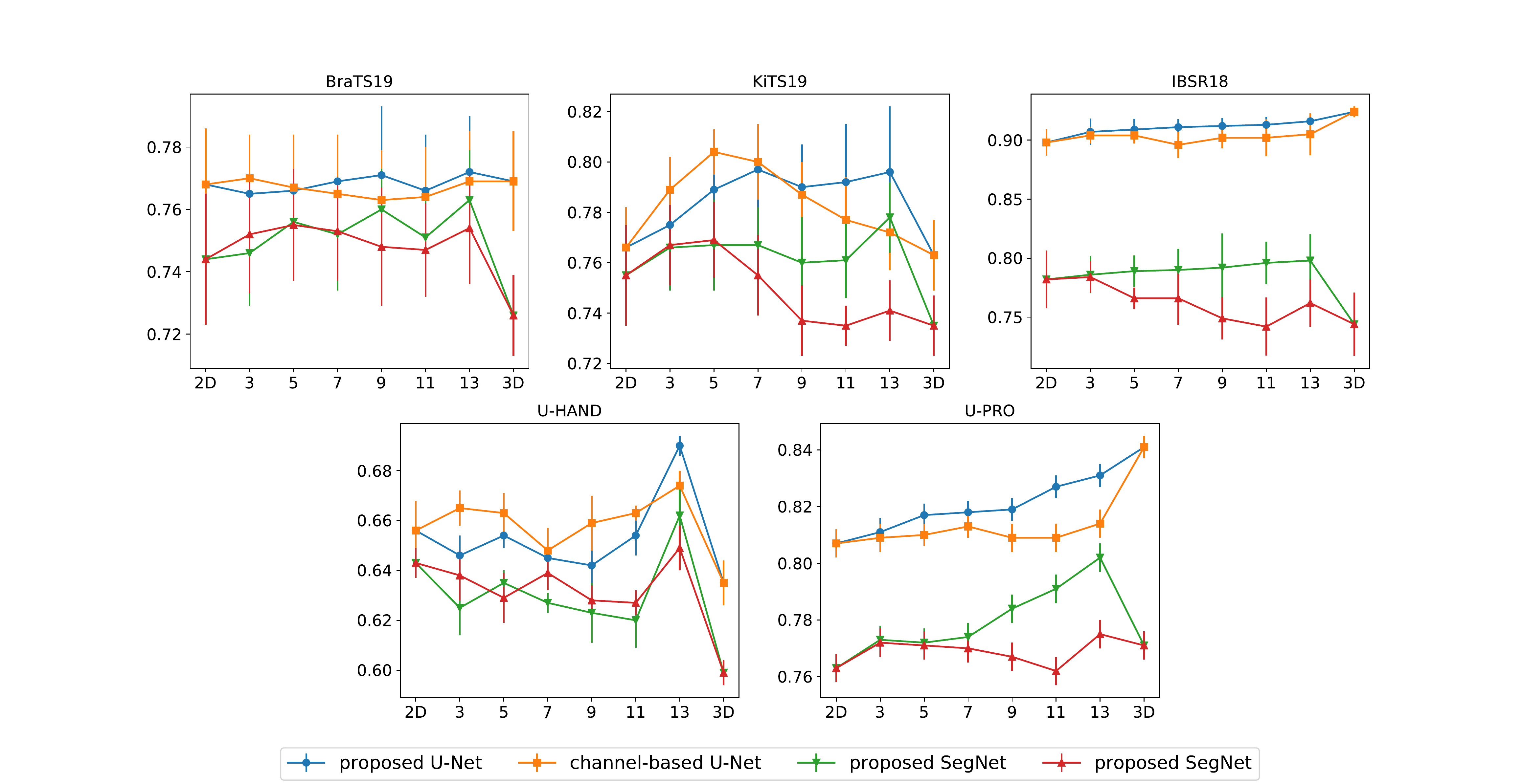}
    \caption{Mean and standard deviation of 5 runs on BraTS19, KiTS19, IBSR18, U-HAND and U-PRO data sets.}    
    \label{fig:errorbar_grid}    
\end{figure*}

The segmentation performances in terms of \gls{dsc} of all models are illustrated in \figref{fig:errorbar_grid}. For each data set, the mean \gls{dsc} scores are plotted (with point-wise standard deviation bars) as a function of the input size, and are given for the 2D, pseudo-3D with $d=3$ to $d=13$, and 3D models, and for the U-Net and SegNet backbones. These results are also tabled in \tableref{tab:performance_test}, along with summaries of the experiment setups per data set. 

Randomly selected example segmentations are illustrated in \figref{fig:qualitative}. For each data set, a prediction from the 2D models, pseudo-3D models with the proposed method, and 3D models are given, along with their respective ground truths. The \gls{brats}, \gls{kits} and \gls{pros} segmentations are cropped for ease of viewing. We chose to omit examples for the channel-based pseudo-3D models because of their high level of similarity to the proposed method. Segmentations with the channel-based method, along with additional exemplary segmentations, can be found in \figref{fig:qualitative_channel_1}-\ref{fig:qualitative_channel_2} in Section~F of the Supplementary Material. 

The computational costs of the models used for \gls{brats} experiments are presented in \tableref{tab:comparison}. The number of model parameters, graphical memory use, and \gls{flops} are only dependent on the model type, and therefore the corresponding columns in \tableref{tab:comparison} are equal for all other data sets. The same variables are shown for the other data sets in \tableref{tab:comparison_kits}--\ref{tab:comparison_pros} in Section~B of the Supplementary Material, where the only differences are in the training and inference times due to the different numbers of samples; these two parameters scale with the data set size.

\section{Discussion}
\label{sec:discussion}

This study evaluated the inclusion of neighboring spatial context as an input of \glspl{cnn} for medical image segmentation. Such pseudo-3D methods with a multi-slice input and single-slice output are commonly implemented by regarding the adjacent slices as additional channels of the central slice. Apart from this approach, we also proposed an alternative pseudo-3D method, based upon multiple preliminary 3D convolution steps before processing by a 2D \gls{cnn}. Across five different data sets and using U-Net and SegNet \gls{cnn} backbones, we compared both these pseudo-3D methods, for an input size $d=3$ up to $d=13$, to end-to-end 2D and 3D \glspl{cnn} with respectively single slice and whole volume inputs and outputs. Additionally, we evaluated a number of computational parameters to get a sense of each model's hardware requirement and load.

\subsection{Computational costs}

As seen in \tableref{tab:comparison}, the computational costs are in line with what would be expected. The transition block adds a relatively small amount of extra parameters on top of the main 2D network, and the required amount of GPU memory and \gls{flops} scale accordingly with $d$. Since the input is still the same size as for the channel-based method, the training times per epoch are largely similar. One advantage of the fully 3D \glspl{cnn} demonstrated in these results, is that prediction time is significantly faster because samples can be processed all at once instead of slice by slice. 


The high computational cost of end-to-end 3D convolution is also demonstrated in \tableref{tab:comparison}. The memory footprint is almost $35$ times larger than the 2D U-Net; over $16$~GB is required to train on the complete volumes, which is at or above the limit of most modern commodity GPUs. Both pseudo-3D methods use less than 5~\% of the GPU memory consumed by the end-to-end 3D network, even at $d=13$. It can thus be concluded that both pseudo-3D methods are computationally very efficient ways of including more inter-slice information, with the proposed method being slightly more expensive in terms of the GPU memory consumption compared to the channel-based method.





\subsection{Quantitative analysis}

As can be seen in \figref{fig:errorbar_grid}, overall, all experiments managed to produce acceptable segmentation results, even for data sets with complex structures such as the \gls{brats} images, or with organs that can be hard to visually distinguish, such as in the \gls{hene} set. One obvious similarity between these data sets is that using a U-Net backbone outperforms the SegNet in nearly every case. Regarding the behavior as a function of input size $d$, the results in \figref{fig:errorbar_grid} are inconclusive for almost all data sets.

In the plots from the \gls{brats}, \gls{kits}, \gls{ibsr} and \gls{hene} data in \figref{fig:errorbar_grid}, there does not seem to be an additional benefit by adding more slices as input over an end-to-end 2D approach. There seem to be some exceptions, like the surge at $d=13$ in the \gls{hene} results, but in these four data sets the variance is either too high or the rate of increase is too low to draw any strong conclusions. For these cases, it would be doubtful if the accessory downsides, \textit{e.g.}~increased training time, are worth the at most marginal improvements in segmentation performance. Likewise, there seems to be no significant difference between our proposed method and the channel-based method in these four data sets.

The only data set in this study where \gls{dsc} does seem to significantly improve with $d$ is in the \gls{pros}. As more slices are being included in the input volume, the segmentation performance approaches that of a fully 3D network, and the proposed method outperforms the channel-based method by an increasing margin. While the overall improvement when going from 2D to pseudo-3D with $d=13$ is arguably low, we can regard the \gls{pros} case as a demonstration of the possibility that pseudo-3D models can improve the segmentation performance over 2D methods.

Fully 3D \glspl{cnn} seems to produce equal or worse results than their 2D and pseudo-3D counterparts in most cases. Again, the only exception seems to be in the \gls{pros} results, and in this case only when the U-Net is used as backbone network (see the respective plot in \figref{fig:errorbar_grid}). This could be explained by the much higher number of parameters of 3D \glspl{cnn}, which makes them prone to overfitting. The high number of data samples in the \gls{pros} set, combined with the skip-connections that differentiates the U-Net from the SegNet, might have been enough to overcome the problem in this specific case.

There does not seem to be a straight-forward explanation as to why the \gls{pros} data set is an exception compared to the other data sets. In an attempt to connect \gls{dsc} behaviour with $d$ to differences in data set properties a feature-based regression analysis was performed. We computed features of the structures (ground truth masks) that describe each mask's structural properties: structure depth (\textit{i.e.}~the average number of consecutive slices a structure is present in), structure size relative to the total volume, and average structural inter-slice spatial displacement. The extracted feature values for all data sets and their respective structures can be found in the Supplementary Material \tableref{tab:structure_brats}--\ref{tab:structure_pros}. We found no significant agreements between models that could connect one of these data set features to \gls{dsc} with respect to $d$. For more details about the feature extraction and regression analysis, see Sections~A.1 and A.2 of the Supplementary Material.

Another distinction between the \gls{pros} set and the others included in this study is its much larger number of samples. As mentioned above, this could have been a contributing factor to the higher performance of the \gls{pros} 3D U-Net compared to other data sets' 3D \gls{cnn} results. This feature was also hypothesized to influence the relation between $d$ and \gls{dsc}, and therefore the following analysis was performed: the same experiments were performed but now training on distinct subsets of $200$ samples from the \gls{pros} data set. The average scores obtained from the five distinct subsets can be found in \figref{fig:pros_200} in the Supplementary Material, where we see a similar behavior as in \figref{fig:errorbar_grid}. Hence, we rule out the data set size as the main cause as well.

We, therefore, conclude that pseudo-3D methods have the potential to increase segmentation performance, but in the general case will not yield better results compared to conventional, end-to-end 2D and 3D \glspl{cnn}. 

Further analysis to explain the behaviour of the \gls{pros} data set might be performed. The regression-based feature analysis was somewhat rudimentary, and could likely be extended with \textit{e.g.} more sophisticated models and more data.

Another possible follow-up study might be to investigate whether it is the multi-slice output (\textit{e.g.}~producing segmentations for all input slices) in pseudo-3D methods that improve the results in other studies. While this was out of the scope of this work, aggregating multiple outputs may be the main reason why pseudo-3D methods sometimes improve the segmentation performance. Based on our conclusions that using multi-slice inputs does not seem to improve the results on their own, the added benefit might only come into play from aggregation of multiple outputs. In this case, using something like Bayesian dropout could prove just as beneficial.

\subsection{Qualitative analysis}

\begin{figure*}[!ht] 
    \centering
    \includegraphics[width=\textwidth]{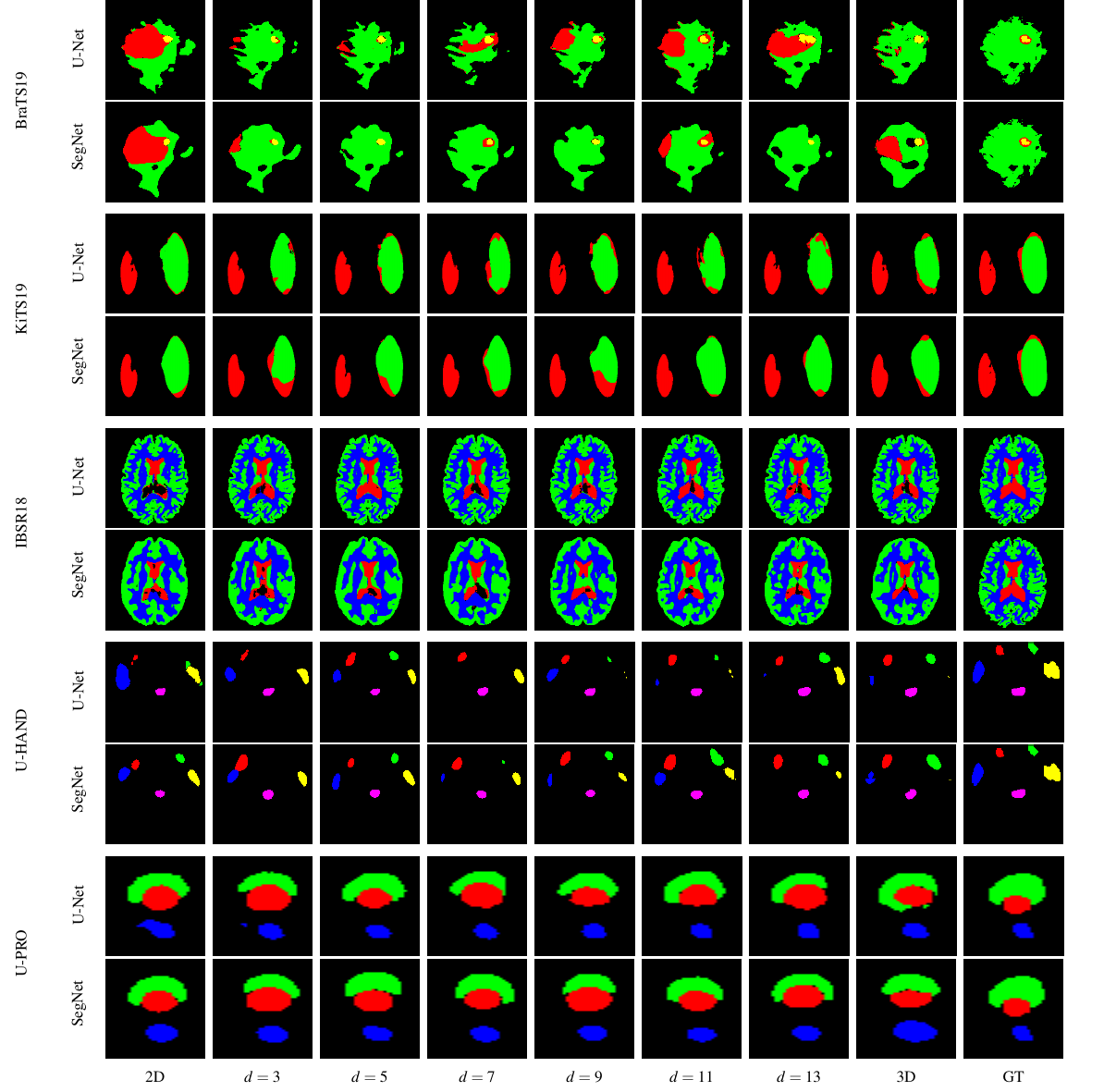}
    \caption{Qualitative results of proposed method on all data sets (see Figure 3 in the Supplementary Material for the qualitative results of the proposed method in the same evaluated examples). From top to bottom: (i) \gls{brats} tumor structures: the necrotic and non-enhancing tumor core (NCR/NET---label 1, red), the peritumoral edema (ED---label 2, green) and the GD-enhancing tumor (ET---label 4, yellow); (ii) \gls{kits} class structure: the kidney (red) and kidney tumor (green); (iii) \gls{ibsr} class structure: cerebrospinal fluid (red), white matter (green) and gray matter (blue); (iv) \gls{hene} class structure: left and right submandibular glands (red and green), left and right parotid glands (dark blue and yellow), larynx (light blue), and medulla oblongata (pink); (v) \gls{pros} class structure: prostate (red), bladder (green) and rectum (blue). From left to right: 2D, $d=3,5,7,9,11,13$, 3D, and ground truth (GT).}
    \label{fig:qualitative}
\end{figure*}

It is important to emphasize that the images in \figref{fig:qualitative} are randomly selected single slices from thousands of samples and are therefore presented purely for illustrative purposes, and might not always be a representation of the overall segmentation performance of a particular data set. However, some remarks can be made that can be related to the quantitative results in \figref{fig:errorbar_grid}. The relatively large variance in segmentation performance between experiments of the \gls{brats} data are demonstrated in \figref{fig:qualitative}; as seen, the predictions can differ quite drastically within the same model and with varying $d$. This reflects the \glspl{dsc} of the \gls{brats} set presented in \figref{fig:errorbar_grid}.






It also appears that the U-Net is better at capturing fine structural details, while the SegNet segmentations seems to be coarser and simpler. This becomes particularly noticeable in data sets with complex structures, such as the gray matter-white matter border in the \gls{ibsr} images (\figref{fig:qualitative}). This in turn results in an overall large difference in mean \gls{dsc} between U-Net and SegNet. When the ground truth structures are more coarsely shaped, such as in the \gls{hene} set, the SegNet can keep up much better with the U-Net performance.

\subsection{Effect of the Loss Function}

In an earlier stage of this project, we employed a different experimental setup with a pure \gls{dsc} loss function. However, these initial experiments proved this loss not to be sufficient for all data sets. Particularly the \gls{kits} and \gls{hene} data sets yielded unacceptably unstable results which, even with exactly equal hyperparameters, could either result in fairly accurate segmentations or complete failure.
Investigation of the \glspl{dsc} of individual structures demonstrated that in these failed experiments, multiple structures did not improve beyond a \gls{dsc} on the order of 0.1. After adapting the loss function to include also the \gls{ce} term (see \eqref{eq:final_loss}), the results improved substantially for all data sets. Performance details for each run using the pure \gls{dsc} and final loss function can be seen in \figref{fig:errorbar_grid_dice} and \tableref{tab:performance_test_dice} in Section E of the Supplementary Material.

\section{Conclusion}
\label{sec:conclusion}

This study systematically evaluated pseudo-3D \glspl{cnn}, where a stack of adjacent slices is used as input for a prediction on the central slice. The hypothesis underlying this approach is that the added neighboring spatial information would improve segmentation performance, with only a small amount of added computational cost compared to an end-to-end 2D \gls{cnn}. However, whether or not this is actually a sensible approach had not previously been evaluated in the literature.

Aside from the conventional method, where the multiple slices are input as multiple channels, we introduced here a novel pseudo-3D method where a subvolume is repeatably convolved in 3D to obtain a final 2D feature map. This 2D feature map is then in turn fed into a standard 2D network.

We investigated the segmentation performance in terms of the \gls{dsc} and the computational cost for a large range of input sizes, for the U-Net and SegNet backbone architectures, and for five diverse data sets covering different anatomical regions, imaging modalities, and segmentation tasks. While pseudo-3D networks can have a large input image size and still be computationally less costly than fully 3D \glspl{cnn} by a large factor, a significant improvement from using multiple input slices was only observed for one of the data sets. We also observed no significant improvement of 3D network performance over 2D networks, regardless of data set size. 

Because of ambiguity in the underlying cause of the behavior on the U-PRO data set compared to the results on the other data sets, we conclude that in the general case pseudo-3D approaches appear to not significantly improve segmentation results over 2D methods.

\section*{Acknowledgments}

This research was conducted using the resources of the High Performance Computing Center North (HPC2N) at Ume{\aa} University, Ume{\aa}, Sweden. We are grateful for the financial support obtained from the Cancer Research Fund in Northern Sweden, Karin and Krister Olsson, Ume\aa{} University, The V\"{a}sterbotten regional county, and Vinnova, the Swedish innovation agency.

\bibliographystyle{IEEEtranN}
\bibliography{bib}

\clearpage
\begin{table*}[!th]
\def\width{2.2 cm}
\def\widthdetail{3.3 cm}
\caption{Mean \gls{dsc} (and standard deviation) of five runs on BraTS19, KiTS19, IBSR18, U-HAND and U-PRO data sets. Models were trained using the summation of the soft \gls{dsc} with the \gls{ce} loss.}
\centering
\begin{tabular}{llccccc}
\toprule
 material/data set   &  & \gls{brats}        & \gls{kits}        & \gls{ibsr}        & \gls{hene}        & \gls{pros}  \\ 
 \midrule
\#epochs            &  & 200                & 200               & 200               & 100               & 100               \\
optimizer           &  & Adam               & Adam              & Adam              & Adam              & Adam              \\
learning rate       &  & $1\cdot10^{-4}$    & $1\cdot10^{-4}$   & $1\cdot10^{-4}$   & $1\cdot10^{-4}$   & $1\cdot10^{-4}$   \\
learning rate drop  &  & $2\cdot10^{-1}$    & $2\cdot10^{-1}$   & $2\cdot10^{-1}$   & $2\cdot10^{-1}$   & $2\cdot10^{-1}$   \\
patience            &  & 5                  & 5                 & 10                & 6                 & 5                 \\
early-stopping      &  & 12                 & 12                & 25                & 14                & 11                \\
\midrule
\multicolumn{7}{@{}l}{U-Net~\citep{ronneberger2015u}}  \\
\midrule
2D                      &  & 0.768 (0.018)              & 0.766 (0.016)             & 0.898 (0.005)             & 0.656 (0.012)                & 0.807 (0.005) \\
proposed ($d=3$)        &  & 0.765 (0.017)              & 0.775 (0.014)             & 0.907 (0.005)             & 0.646 (0.008)                & 0.811 (0.005) \\
proposed ($d=5$)        &  & 0.766 (0.018)              & 0.789 (0.014)             & 0.909 (0.004)             & 0.654 (0.005)                & 0.817 (0.004) \\
proposed ($d=7$)        &  & 0.769 (0.013)              & 0.797 (0.017)             & 0.911 (0.003)             & 0.645 (0.012)                & 0.818 (0.004) \\
proposed ($d=9$)        &  & 0.771 (0.022)              & 0.790 (0.017)             & 0.912 (0.003)             & 0.642 (0.007)                & 0.819 (0.004) \\
proposed ($d=11$)       &  & 0.766 (0.018)              & 0.792 (0.023)             & 0.913 (0.003)             & 0.654 (0.008)                & 0.827 (0.004) \\
proposed ($d=13$)       &  & 0.772 (0.018)              & 0.796 (0.026)             & 0.916 (0.003)             & 0.690 (0.004)                & 0.831 (0.004) \\
\midrule
channel-based ($d=3$)   &  & 0.770 (0.014)              & 0.789 (0.013)             & 0.904 (0.003)             & 0.665 (0.007)                & 0.809 (0.005) \\
channel-based ($d=5$)   &  & 0.767 (0.017)              & 0.804 (0.009)             & 0.904 (0.003)             & 0.663 (0.008)                & 0.810 (0.004) \\
channel-based ($d=7$)   &  & 0.765 (0.019)              & 0.800 (0.015)             & 0.896 (0.005)             & 0.648 (0.009)                & 0.813 (0.004) \\
channel-based ($d=9$)   &  & 0.763 (0.016)              & 0.787 (0.013)             & 0.902 (0.004)             & 0.659 (0.011)                & 0.809 (0.005) \\
channel-based ($d=11$)  &  & 0.764 (0.016)              & 0.777 (0.017)             & 0.902 (0.007)             & 0.663 (0.003)                & 0.809 (0.005) \\
channel-based ($d=13$)  &  & 0.769 (0.016)              & 0.772 (0.015)             & 0.905 (0.008)             & 0.674 (0.006)                & 0.814 (0.005) \\
3D                      &  & 0.769 (0.016)              & 0.763 (0.014)             & 0.924 (0.002)             & 0.635 (0.009)                & 0.841 (0.004) \\ 
\midrule
\multicolumn{7}{@{}l}{Seg-Net~\citep{badrinarayanan2017segnet}}  \\
\midrule
2D                      &  & 0.744 (0.021)              & 0.755 (0.020)             & 0.782 (0.011)             & 0.643 (0.006)                & 0.763 (0.005) \\
proposed ($d=3$)        &  & 0.746 (0.017)              & 0.766 (0.017)             & 0.786 (0.007)             & 0.625 (0.011)                & 0.773 (0.005) \\
proposed ($d=5$)        &  & 0.756 (0.017)              & 0.767 (0.018)             & 0.789 (0.006)             & 0.635 (0.005)                & 0.772 (0.005) \\
proposed ($d=7$)        &  & 0.752 (0.018)              & 0.767 (0.015)             & 0.790 (0.008)             & 0.627 (0.004)                & 0.774 (0.005) \\
proposed ($d=9$)        &  & 0.760 (0.013)              & 0.760 (0.018)             & 0.792 (0.013)             & 0.623 (0.012)                & 0.784 (0.005) \\
proposed ($d=11$)       &  & 0.751 (0.016)              & 0.761 (0.015)             & 0.796 (0.008)             & 0.620 (0.011)                & 0.791 (0.005) \\
proposed ($d=13$)       &  & 0.763 (0.016)              & 0.778 (0.014)             & 0.798 (0.010)             & 0.662 (0.011)                & 0.802 (0.005) \\
\midrule
channel-based ($d=3$)   &  & 0.752 (0.019)              & 0.767 (0.016)             & 0.784 (0.006)             & 0.638 (0.010)                & 0.772 (0.005) \\
channel-based ($d=5$)   &  & 0.755 (0.018)              & 0.769 (0.015)             & 0.766 (0.004)             & 0.629 (0.010)                & 0.771 (0.005) \\
channel-based ($d=7$)   &  & 0.753 (0.016)              & 0.755 (0.016)             & 0.766 (0.010)             & 0.639 (0.007)                & 0.770 (0.005) \\
channel-based ($d=9$)   &  & 0.748 (0.019)              & 0.737 (0.014)             & 0.749 (0.008)             & 0.628 (0.006)                & 0.767 (0.005) \\
channel-based ($d=11$)  &  & 0.747 (0.015)              & 0.735 (0.008)             & 0.742 (0.011)             & 0.627 (0.005)                & 0.762 (0.005) \\
channel-based ($d=13$)  &  & 0.754 (0.018)              & 0.741 (0.012)             & 0.762 (0.009)             & 0.649 (0.009)                & 0.775 (0.005) \\
3D                      &  & 0.726 (0.013)              & 0.735 (0.012)             & 0.744 (0.012)             & 0.599 (0.005)                & 0.771 (0.005) \\
\bottomrule
\end{tabular}
\label{tab:performance_test}
\end{table*}

\clearpage

\clearpage

\section*{Supplementary Material}
\subsection{Structure Analysis}
\subsubsection{Feature Extraction}
We selected three data set features that describe each data set's structural properties: structure depth, structure size relative to the total volume, and average structural inter-slice spatial displacement (see \tableref{tab:structure_brats}-\ref{tab:structure_pros}). These aforementioned structural properties are computed as follows.


The structure depth of class $c=1,\ldots,C$ is computed as
\begin{equation}
    \Phi_c = \frac{\sum_{p=1}^{P}\sum_{r_{pc}=1}^{R_{pc}}\phi_{r_{pc}}}{P\sum_{r_{pc}=1}^{R_{pc}}r_{pc}},
\end{equation}
where $p=1,\ldots,P$ denotes the patient $p$, $r_{pc}$ represents an unconnected region of class $c$ in patient $p$, the $\phi_{r_{pc}}$ denotes the number of consecutive slices (in the axial dimension) of region $r_{pc}$. Here $P$ and $R_{pc}$ are the number of patients and unconnected regions of class $c$ in patient $p$, respectively.

The structure size relative to the total volume of class $c$ is defined as 
\begin{equation}
    \Upsilon_c = \frac{\sum_{p=1}^{P}\upsilon_{cp}}{P \cdot H \cdot W \cdot D},
\end{equation}
where $\upsilon_{cp}$ denotes the total number of voxels labeled as class $c$ in patient $p$. The $H$, $W$, and $D$ are the height, width, and depth of the input volume, respectively.

To compute the structure spatial displacement, we first compute the center of mass, $\psi_{cps}$, of class $c$ of patient $p$ at slice $s$ (in the axial dimension) as
\begin{equation}
    \psi_{cps} = \frac{1}{\upsilon_{cps}} \sum_{i=1}^{H} \sum_{j=1}^{W} r_{ijcps},
\end{equation}
where $r_{ijcps}$ is the value of a voxel at coordinates $(i,j)$ in class $c$ in patient $p$ and slice $s$, and where $\upsilon_{cps}$ denotes the total number of voxels labeled as class $c$ in patient $p$ in slice $s$.

With these, the structure spatial displacement of class $c$ is computed as
\begin{equation}
    \Psi_{c} = \frac{1}{P \cdot D} \sum_{p=1}^{P} \sum_{s=2}^{D} \left\| \psi_{cp(s-1)}, \psi_{cps}\right\|_2,
\end{equation}

where $\| \psi_{cp(s-1)}, \psi_{cps} \|_2$ denotes the Euclidian distance between two coordinate points, $\psi_{cp(s-1)}$ and $\psi_{cps}$.

\subsubsection{Regression Analysis}

To find explanations to why the \acrshort{pros} data set is an exception compared to the other data sets, we attempted to connect \acrshort{dsc} behaviour with $d$, that is the number of slices extracted from the whole volume with a total of $D$ slices as the subvolume input, to differences in data set properties including structure depth, structure size relative to the total volume, and average structural inter-slice spatial displacement, and the number of training samples.

We aggregated all these structure properties to generate minimum, mean and maximum of $\Phi$, $\Upsilon$ and $\Psi$ over all classes in each data set, such that for each data set we obtained a fixed set of nine features. Overall, we formed a regression task to compute regression models for all models (U-Net and SegNet), architectures (2D, 3D, proposed, and channel-based), $d$s (number of slices extracted from the whole volume), and data sets including \gls{brats}, \gls{kits}, \gls{ibsr}, \gls{hene} and \gls{pros} with the following input features:
\begin{itemize}[noitemsep]
    \item $\Phi_{\text{min}}$, minimum of structure depth over classes,
    \item $\Phi_{\text{mean}}$, average of structure depth over classes,
    \item $\Phi_{\text{max}}$, maximum of structure depth over classes,
    \item $\Upsilon_{\text{min}}$, minimum of structure size relative to the total volume over classes,
    \item $\Upsilon_{\text{mean}}$, mean of structure size relative to the total volume over classes,
    \item $\Upsilon_{\text{max}}$, maximum of structure size relative to the total volume over classes,
    \item $\Psi_{\text{min}}$, minimum of spatial displacement over classes,
    \item $\Psi_{\text{mean}}$, average of spatial displacement over classes,
    \item $\Psi_{\text{max}}$, maximum of spatial displacement over classes.
\end{itemize}

We used the Bootstrap (with $1\,000$ rounds) to compute the mean regression coefficient vectors and the corresponding confidence intervals the some regularised linear regression models. In the regression analysis, we used Ridge regression, Lasso, Elastic Net, and Bayesian ARD regression. The analysis was performed using scikit-learn 0.22\footnote{\url{https://scikit-learn.org/stable/}}. However, the regression analysis was non-conclusive, revealing no relation between the structure features and the number of input slices, $d$.


\subsubsection{Extracted Features} See \tableref{tab:structure_brats}-\ref{tab:structure_pros} for more details.

\begin{table*}[!ht]
\def\widthdetail{4.5cm}
\def\width{2.5cm}
\caption{Mean and standard deviation of of each class's depth, $\Phi_c$, count over the total number of pixels, $\Upsilon_c$, spatial displacement, $\Psi_{c}$, in the axial direction over all volumes in the BraTS19 data set.}
\centering
\begin{adjustbox}{max width=\textwidth}
\begin{tabular}{L{\widthdetail}lC{\width}C{\width}C{\width}}
\toprule
class  &            & $\Phi_c$              & $\Upsilon_c$          & $\Psi_{c}$        \\ 
\midrule
non-enhancing tumor core             
                    &  & 33.3 (17.20)       & 25.20 (33.70)         & 1.76 (0.94)       \\
peritumoral edema   &  & 60.6 (16.50)       & 61.90 (47.60)         & 1.65 (0.85)       \\
GD-enhancing tumor  &  & 31.9 (18.10)       & 18.80 (20.50)         & 1.36 (1.99)       \\
\bottomrule
\end{tabular}
\end{adjustbox}
\label{tab:structure_brats}
\end{table*}

\begin{table*}[!ht]
\def\widthdetail{4.5cm}
\def\width{2.5cm}
\caption{Mean and standard deviation of of each class's depth, $\Phi_c$, count over the total number of pixels, $\Upsilon_c$, spatial displacement, $\Psi_{c}$, in the axial direction over all volumes in the KiTS19 data set.}
\centering
\begin{adjustbox}{max width=\textwidth}
\begin{tabular}{L{\widthdetail}lC{\width}C{\width}C{\width}}
\toprule
class  &            & $\Phi_c$              & $\Upsilon_c$          & $\Psi_{c}$        \\ 
\midrule
kidney              &  & 37.34 (4.34)       & 182.20 (50.00)        & 4.12 (0.98)       \\
kidney tumor        &  & 15.64 (10.84)      & 53.49 (107.85)        & 1.90 (1.39)       \\
\bottomrule
\end{tabular}
\end{adjustbox}
\label{tab:structure_kits}
\end{table*}

\begin{table*}[!ht]
\def\widthdetail{4.5cm}
\def\width{2.5cm}
\caption{Mean and standard deviation of of each class's depth, $\Phi_c$, count over the total number of pixels, $\Upsilon_c$, spatial displacement, $\Psi_{c}$, in the axial direction over all volumes in the IBSR18 data set.}
\centering
\begin{adjustbox}{max width=\textwidth}
\begin{tabular}{L{\widthdetail}lC{\width}C{\width}C{\width}}
\toprule
class  &            & $\Phi_c$              & $\Upsilon_c$          & $\Psi_{c}$        \\ 
\midrule
cerebrospinal fluid &  & 55.67 (13.42)      & 18.31 (7.56)          & 1.89 (0.29)       \\
white matter        &  & 140.06 (12.55)     & 815.62 (205.51)       & 0.82 (0.79)       \\
gray matter         &  & 114.78 (13.01)     & 431.49 (77.95)        & 0.98 (0.11)       \\
\bottomrule
\end{tabular}
\end{adjustbox}
\label{tab:structure_ibsr}
\end{table*}

\begin{table*}[!ht]
\def\widthdetail{4.5cm}
\def\width{2.5cm}
\caption{Mean and standard deviation of of each class's depth, $\Phi_c$, count over the total number of pixels, $\Upsilon_c$, spatial displacement, $\Psi_{c}$, in the axial direction over all volumes in the U-HAND data set.}
\centering
\begin{adjustbox}{max width=\textwidth}
\begin{tabular}{L{\widthdetail}lC{\width}C{\width}C{\width}}
\toprule
class  &            & $\Phi_c$              & $\Upsilon_c$          & $\Psi_{c}$        \\ 
\midrule
left submandibular glands 
                    &  & 5.19 (1.29)        & 2.38 (0.81)           & 1.67 (0.66)       \\
right submandibular glands 
                    &  & 5.21 (1.43)        & 2.34 (0.78)           & 1.87 (0.82)       \\
left parotid glands &  & 7.53 (2.09)        & 7.16 (2.96)           & 1.99 (0.82)       \\
right parotid glands 
                    &  & 7.41 (2.15)        & 6.89 (2.83)           & 2.01 (0.65)       \\
larynx              &  & 5.38 (1.31)        & 10.62 (3.35)          & 1.45 (0.74)       \\
medulla oblongata   &  & 30.81 (5.35)       & 11.52 (2.88)          & 2.01 (0.54)       \\
\bottomrule
\end{tabular}
\end{adjustbox}
\label{tab:structure_hene}
\end{table*}

\begin{table*}[!ht]
\def\widthdetail{4.5cm}
\def\width{2.5cm}
\caption{Mean and standard deviation of of each class's depth, $\Phi_c$, count over the total number of pixels, $\Upsilon_c$, spatial displacement, $\Psi_{c}$, in the axial direction over all volumes in the U-PRO data set.}
\centering
\begin{adjustbox}{max width=\textwidth}
\begin{tabular}{L{\widthdetail}lC{\width}C{\width}C{\width}}
\toprule
class  &            & $\Phi_c$              & $\Upsilon_c$          & $\Psi_{c}$        \\ 
\midrule
prostate            &  & 10.7 (2.30)        & 2.34 (1.02)           & 0.63 (0.25)       \\
bladder             &  & 10.2 (3.90)        & 4.80 (3.18)           & 1.45 (0.69)       \\
rectum              &  & 25.0 (3.50)        & 2.99 (1.10)           & 0.95 (0.26)       \\
\bottomrule
\end{tabular}
\end{adjustbox}
\label{tab:structure_pros}
\end{table*}

\subsection{Supplementary Computational Results}

To compare the computational cost of our proposed models to the corresponding 2D and 3D \gls{cnn} models, we extracted the number of trainable parameters, the maximum amount of \gls{gpu} memory used, the number of \gls{flops}, training time per epoch, and prediction time per sample.

The computational costs of the models used for \gls{brats} experiments are presented in Table~2 of the main paper. The number of model parameters, graphical memory use, and \gls{flops} are only dependent on the model type, and therefore are equal for all other data sets. The same variables are shown here for the other data sets in \tableref{tab:comparison_kits}--\ref{tab:comparison_pros}, where the only differences are in the training and inference times due to the different numbers of samples; these two parameters scale with the data set size.

\begin{table*}[!h]
\def\widthdetail{2.6 cm}
\def\width{1.9 cm}
\def\widthnarrow{1.6 cm}
\def\widthnarrowest{1 cm}
\caption{Architecture comparison. Experiment on U-Net architecture on KiTS19 data set. Patch shape was set at $256 \times 256 \times d$ where $d$ is the number of slices. Here, \textit{t} and \textit{p} denote the training time per epoch and prediction time per sample, respectively. This experiment was performed on an NVIDIA GeForce GTX 1080 Ti.}
\centering
\begin{adjustbox}{max width=\textwidth}
\begin{tabular}{llrrrrrr}
\toprule
Model                   &  & \#slices                  
                                    & \#params  & memory    & \#FLOPS           
                                                                        & \textit{t} per epoch 
                                                                                    & \textit{p} per sample \\ 
\midrule
2D                      &  & 1      & 494k      & 468MB     & 2.456M    & 87s       & 19.04s    \\
\midrule
\multirow{6}{*}{proposed}
                        &  & 3      & 497k      & 498MB     & 2.469M    & 118s      & 19.40s    \\
                        &  & 5      & 504k      & 520MB     & 2.504M    & 131s      & 20.10s    \\
                        &  & 7      & 511k      & 542MB     & 2.538M    & 201s      & 21.35s    \\
                        &  & 9      & 518k      & 565MB     & 2.573M    & 255s      & 22.07s    \\
                        &  & 11     & 525k      & 587MB     & 2.608M    & 408s      & 22.69s    \\
                        &  & 13     & 530k      & 609MB     & 2.643M    & 576s      & 23.07s    \\
\midrule
\multirow{6}{*}{channel-based}
                        &  & 3      & 495k      & 486MB     & 2.457M    & 91s       & 19.30s    \\
                        &  & 5      & 495k      & 498MB     & 2.459M    & 93s       & 19.04s    \\
                        &  & 7      & 495k      & 511MB     & 2.460M    & 169s      & 19.57s    \\
                        &  & 9      & 496k      & 524MB     & 2.462M    & 240s      & 19.57s    \\
                        &  & 11     & 496k      & 535MB     & 2.463M    & 395s      & 19.83s    \\
                        &  & 13     & 497k      & 546MB     & 2.465M    & 551s      & 20.01s    \\
\midrule
3D                      &  & 32     & 1\,460k   & 16\,315MB & 7.297M    & 1\,041s   & 3.08 s    \\
\bottomrule
\end{tabular}
\end{adjustbox}
\label{tab:comparison_kits}
\end{table*}

\begin{table*}[!h]
\def\widthdetail{2.6 cm}
\def\width{1.9 cm}
\def\widthnarrow{1.6 cm}
\def\widthnarrowest{1 cm}
\caption{Architecture comparison. Experiment on U-Net architecture on IBSR data set. Patch shape was set at $256 \times 128 \times d$ where $d$ is the number of slices. Here, \textit{t} and \textit{p} denote the training time per epoch and prediction time per sample, respectively. This experiment was performed on an NVIDIA GeForce GTX 1080 Ti.}
\centering
\begin{adjustbox}{max width=\textwidth}
\begin{tabular}{llrrrrrr}
\toprule
Model                   &  & \#slices                  
                                    & \#params  & memory    & \#FLOPS           
                                                                        & \textit{t} per epoch 
                                                                                    & \textit{p} per sample \\ 
\midrule
2D                      &  & 1      & 494       & 457       & 2.398M    & 11s       & 10.02s \\
\midrule
\multirow{6}{*}{proposed}
                        &  & 3      & 497k      & 486MB     & 2.410M    & 15s       & 10.21s \\
                        &  & 5      & 504k      & 507MB     & 2.441M    & 17s       & 10.58s \\
                        &  & 7      & 511k      & 528MB     & 2.472M    & 25s       & 11.24s \\
                        &  & 9      & 518k      & 550MB     & 2.505M    & 32s       & 11.62s \\
                        &  & 11     & 525k      & 571MB     & 2.537M    & 52s       & 11.95s \\
                        &  & 13     & 530k      & 592MB     & 2.569M    & 73s       & 12.15s \\
\midrule
\multirow{6}{*}{channel-based}
                        &  & 3      & 495k      & 472MB     & 2.386M    & 12s       & 10.16s \\
                        &  & 5      & 495k      & 484MB     & 2.390M    & 12s       & 10.02s \\
                        &  & 7      & 495k      & 497MB     & 2.393M    & 21s       & 10.30s \\
                        &  & 9      & 496k      & 510MB     & 2.396M    & 30s       & 10.30s \\
                        &  & 11     & 496k      & 521MB     & 2.399M    & 50s       & 10.44s \\
                        &  & 13     & 497k      & 532MB     & 2.402M    & 70s       & 10.53s \\
\midrule
3D                      &  & 32     & 1\,460k   & 15\,897MB & 7.11M     & 132s      & 1.62s  \\ 
\bottomrule
\end{tabular}
\end{adjustbox}
\label{tab:comparison_ibsr}
\end{table*}

\begin{table*}[!h]
\def\widthdetail{2.6 cm}
\def\width{1.9 cm}
\def\widthnarrow{1.6 cm}
\def\widthnarrowest{1 cm}
\caption{Architecture comparison. Experiment on U-Net architecture on U-HAND data set. Patch shape was set at $256 \times 256 \times d$ where $d$ is the number of slices. Here, \textit{t} and \textit{p} denote the training time per epoch and prediction time per sample, respectively. This experiment was performed on an NVIDIA GeForce GTX 1080 Ti.}
\centering
\begin{adjustbox}{max width=\textwidth}
\begin{tabular}{llrrrrrr}
\toprule
Model                   &  & \#slices                  
                                    & \#params  & memory    & \#FLOPS           
                                                                        & \textit{t} per epoch 
                                                                                    & \textit{p} per sample \\ 
\midrule
2D                      &  & 1      & 494k      & 468MB     & 2.456M    & 16s       & 19.04s    \\
\midrule
\multirow{6}{*}{proposed}
                        &  & 3      & 497k      & 498MB     & 2.469M    & 22s       & 19.40s    \\
                        &  & 5      & 504k      & 520MB     & 2.504M    & 24s       & 20.10s    \\
                        &  & 7      & 511k      & 542MB     & 2.538M    & 37s       & 21.35s    \\
                        &  & 9      & 518k      & 565MB     & 2.573M    & 47s       & 22.07s    \\
                        &  & 11     & 525k      & 587MB     & 2.608M    & 75s       & 22.69s    \\
                        &  & 13     & 530k      & 609MB     & 2.643M    & 106s      & 23.07s    \\
\midrule
\multirow{6}{*}{channel-based}
                        &  & 3      & 495k      & 486MB     & 2.457M    & 17s       & 19.30s    \\
                        &  & 5      & 495k      & 498MB     & 2.459M    & 17s       & 19.04s    \\
                        &  & 7      & 495k      & 511MB     & 2.460M    & 31s       & 19.57s    \\
                        &  & 9      & 496k      & 524MB     & 2.462M    & 44s       & 19.57s    \\
                        &  & 11     & 496k      & 535MB     & 2.463M    & 73s       & 19.83s    \\
                        &  & 13     & 497k      & 546MB     & 2.465M    & 101       & 20.01s    \\
\midrule
3D                      &  & 32     & 1\,460k   & 16\,315MB & 7.297M    & 191       & 3.08 s    \\
\bottomrule
\end{tabular}
\end{adjustbox}
\label{tab:comparison_hene}
\end{table*}

\begin{table*}[!h]
\def\widthdetail{2.6 cm}
\def\width{1.9 cm}
\def\widthnarrow{1.6 cm}
\def\widthnarrowest{1 cm}
\caption{Architecture comparison. Experiment on U-Net architecture on U-PRO data set. Patch shape was set at $256 \times 256 \times d$ where $d$ is the number of slices. Here, \textit{t} and \textit{p} denote the training time per epoch and prediction time per sample, respectively. This experiment was performed on an NVIDIA GeForce GTX 1080 Ti.}
\centering
\begin{adjustbox}{max width=\textwidth}
\begin{tabular}{llrrrrrr}
\toprule
Model                   &  & \#slices                  
                                    & \#params  & memory    & \#FLOPS           
                                                                        & \textit{t} per epoch 
                                                                                    & \textit{p} per sample \\ 
\midrule
2D                      &  & 1      & 494k      & 468MB     & 2.456M    & 250s      & 19.04s    \\
\midrule
\multirow{6}{*}{proposed}
                        &  & 3      & 497k      & 498MB     & 2.469M    & 340s      & 19.40s    \\
                        &  & 5      & 504k      & 520MB     & 2.504M    & 376s      & 20.10s    \\
                        &  & 7      & 511k      & 542MB     & 2.538M    & 578s      & 21.35s    \\
                        &  & 9      & 518k      & 565MB     & 2.573M    & 732s      & 22.07s    \\
                        &  & 11     & 525k      & 587MB     & 2.608M    & 1\,171s   & 22.69s    \\
                        &  & 13     & 530k      & 609MB     & 2.643M    & 1\,654s   & 23.07s    \\
\midrule
\multirow{6}{*}{channel-based}
                        &  & 3      & 495k      & 486MB     & 2.457M    & 262s      & 19.30s    \\
                        &  & 5      & 495k      & 498MB     & 2.459M    & 268s      & 19.04s    \\
                        &  & 7      & 495k      & 511MB     & 2.460M    & 485s      & 19.57s    \\
                        &  & 9      & 496k      & 524MB     & 2.462M    & 689s      & 19.57s    \\
                        &  & 11     & 496k      & 535MB     & 2.463M    & 1\,136s   & 19.83s    \\
                        &  & 13     & 497k      & 546MB     & 2.465M    & 1\,583s   & 20.01s    \\
\midrule
3D                      &  & 32     & 1\,460k   & 16\,315MB & 7.297M    & 2\,990s   & 3.08 s    \\
\bottomrule
\end{tabular}
\end{adjustbox}
\label{tab:comparison_pros}
\end{table*}

\subsection{SegNet Pseudo-3D architecture}
\begin{figure*}[!th] 
\centering
\includegraphics[width=0.8\textwidth]{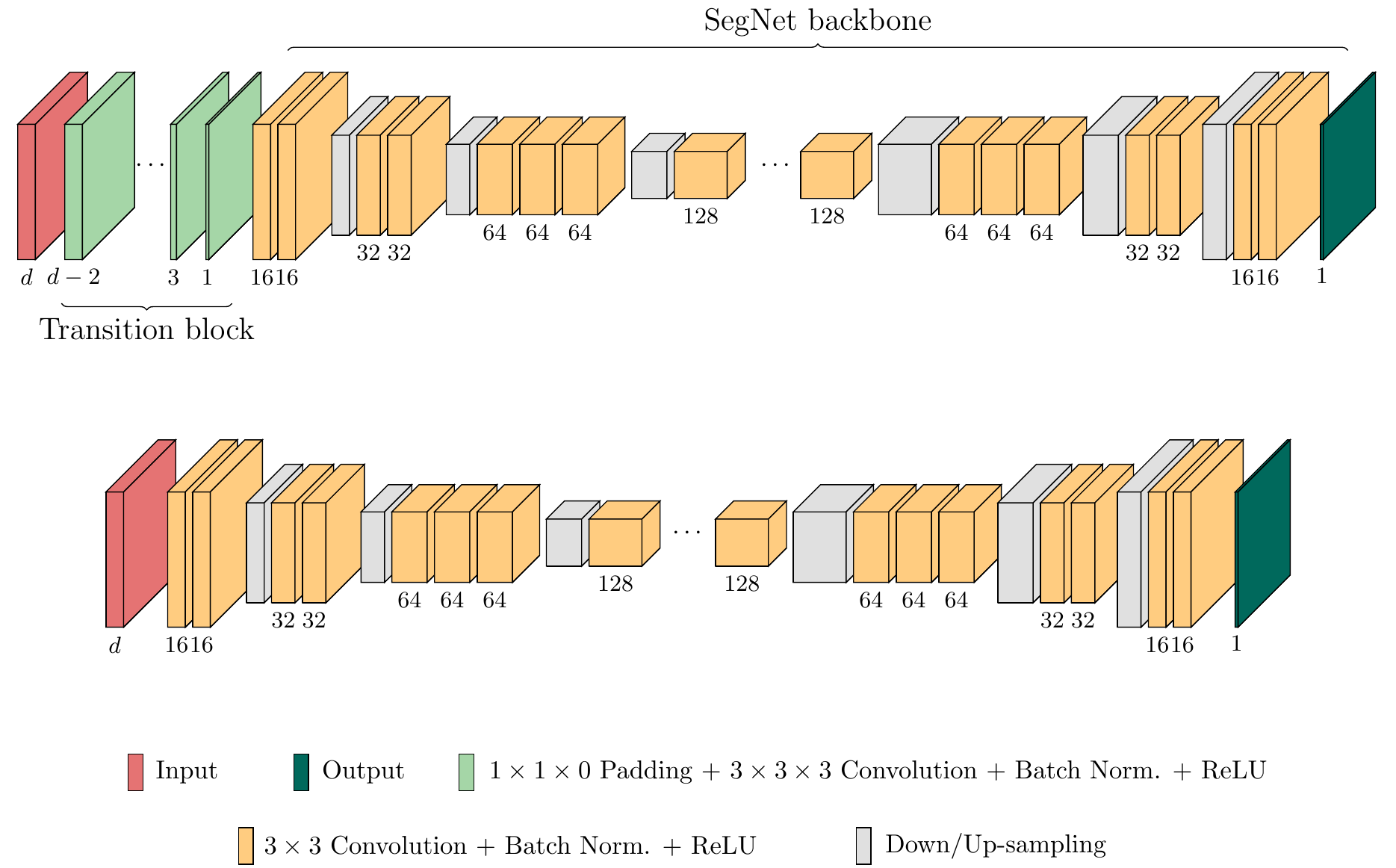}
\caption{Proposed methods with a SegNet backbone. The output is the prediction for the central slice of the input. The numbers in the transition block indicate the depth and in the backbone the number of filters. Top: a transition block uses 3D convolution and 2D padding to iteratively reduce the input from depth $d$ to $1$, while the width and height do not change. Bottom: the neighboring slices are regarded as multiple channels, and the input can be fed into the 2D \acrshort{cnn} right away.
}
\label{fig:proposed_seg}
\end{figure*}

Figure~2 in the main paper shows both pseudo-3D methods with a U-Net backbone. Here, \figref{fig:proposed_seg} shows the same methods but with a SegNet backbone.

\subsection{\acrshort{pros} Subset Experiments Results}
\begin{figure*}[!h]
    \centering
    \includegraphics[width=.4\textwidth]{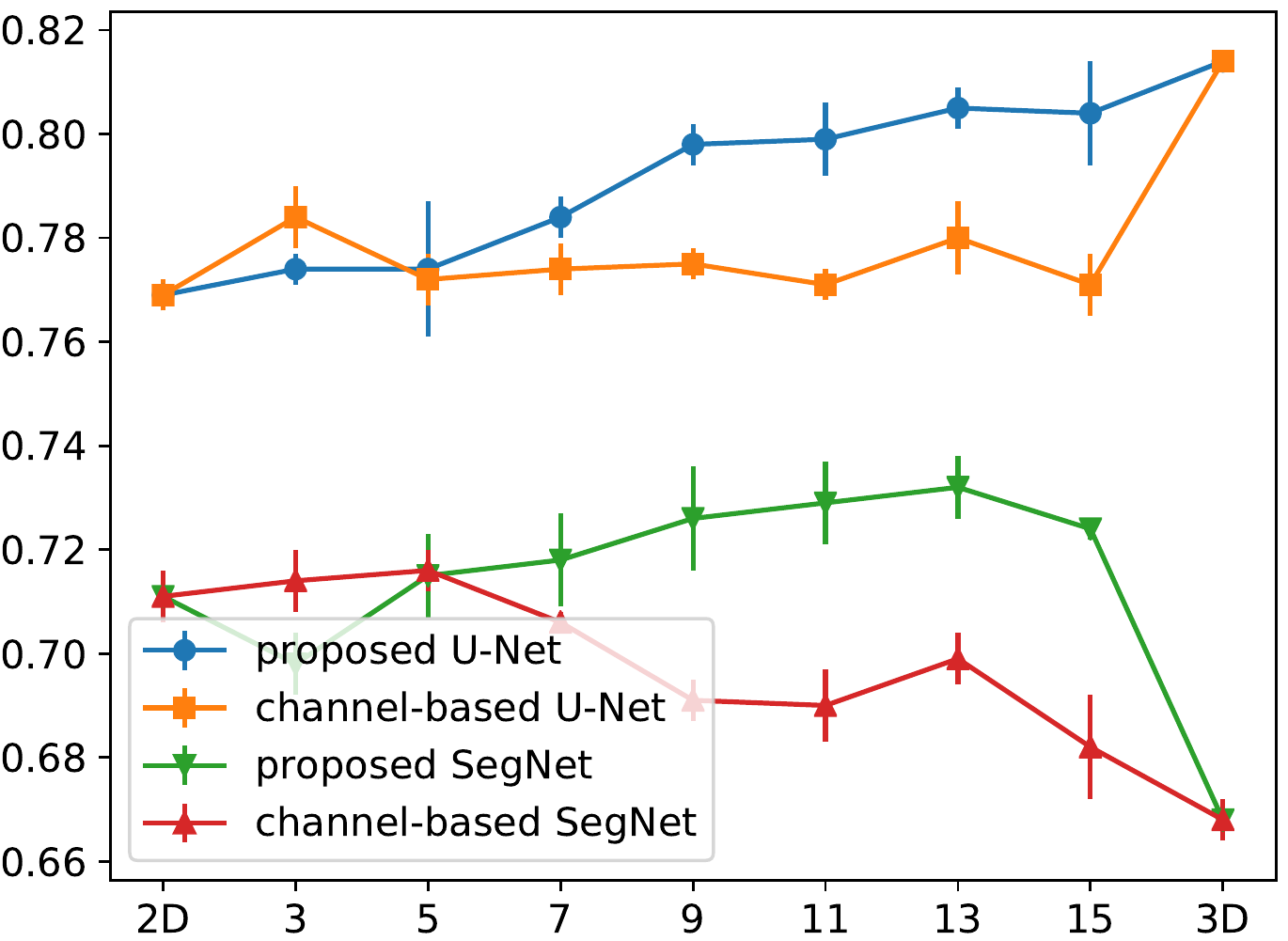}
    \caption{Mean and standard error of test set on the U-PRO data set. Each run was trained on 200 patients.}   
    \label{fig:pros_200}
\end{figure*}

A distinction between the \gls{pros} set and the others included in this study is its much larger number of samples. This feature was hypothesized to influence the relation between $d$ and \gls{dsc}, and therefore the following analysis was performed: the same experiments were performed but now training on distinct subsets of $200$ samples from the \gls{pros} data set. The average scores obtained from the five distinct subsets can be found here in \figref{fig:pros_200}, where we see a similar behavior as in Figure~9 in the main paper. Hence, we rule out the data set size as the main cause of the \gls{pros} performance behaviour.

\subsection{Supplementary Quantitative Results}

\begin{figure*}[!th]
    \centering
    \includegraphics[trim=120 0 120 40, clip, width=\textwidth]{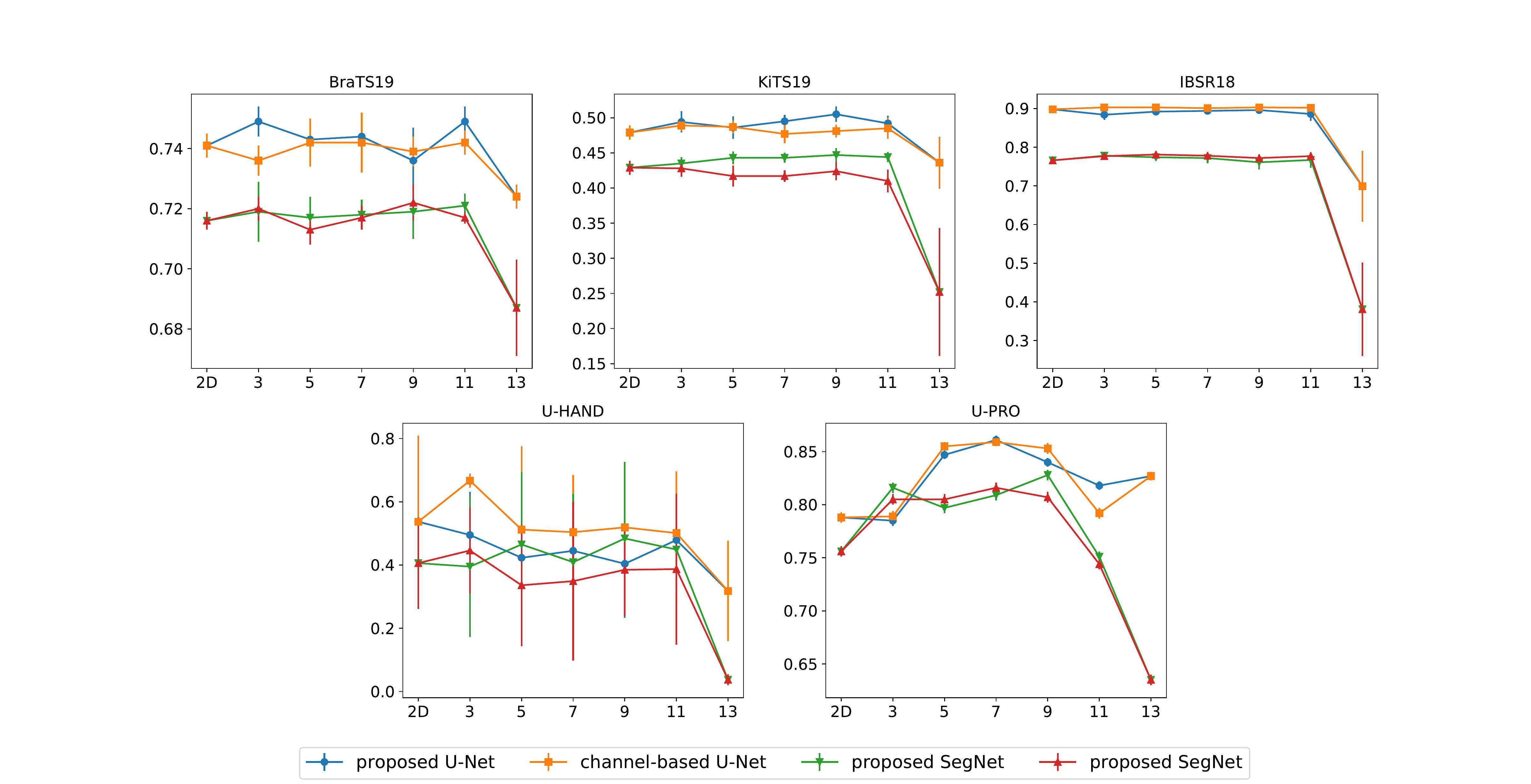}
    \caption{Mean and standard deviation of five runs on BraTS19, KiTS19, IBSR18, U-HAND and U-PRO data sets using the soft \gls{dsc} loss.}
    \label{fig:errorbar_grid_dice}    
\end{figure*}

In an earlier stage of this project, we employed a different experimental setup with a pure \gls{dsc} loss function. However, these initial experiments proved this loss not to be sufficient for all data sets. Particularly the \gls{kits} and \gls{hene} data sets yielded unacceptably unstable results which, even with exactly equal hyperparameters, could either result in fairly accurate segmentations or complete failure.
Investigation of the \glspl{dsc} of individual structures demonstrated that in these failed experiments, multiple structures did not improve beyond a \gls{dsc} on the order of 0.1. After adapting the loss function to include also the \gls{ce} term, the results improved substantially for all data sets. Performance details for each run using the pure \gls{dsc} and final loss function can be seen in \figref{fig:errorbar_grid_dice} and \tableref{tab:performance_test}.

\begin{table*}[!th]
\caption{Mean \gls{dsc} (and standard deviation) of five runs on BraTS19, KiTS19, IBSR18, U-HAND and U-PRO data sets. Models were trained using the soft \gls{dsc} loss.}
\centering
\begin{adjustbox}{max width=1.1\textwidth}
\hspace{-1.5cm}
\begin{tabular}{llccccc}
\toprule
 material/data set      &  & \acrshort{brats}           & \acrshort{kits}           & \acrshort{ibsr}           & \acrshort{hene}               & \acrshort{pros}  \\ 
 \midrule
\#epochs                &  & 200                & 200               & 200               & 100               & 100               \\
optimizer               &  & Adam               & Adam              & Adam              & Adam              & Adam              \\
learning rate           &  & $1\cdot10^{-4}$    & $1\cdot10^{-4}$   & $1\cdot10^{-4}$   & $1\cdot10^{-4}$   & $1\cdot10^{-4}$   \\
learning rate drop      &  & $2\cdot10^{-1}$    & $2\cdot10^{-1}$   & $2\cdot10^{-1}$   & $2\cdot10^{-1}$   & $2\cdot10^{-1}$   \\
patience                &  & 5                  & 5                 & 10                & 6                 & 5                 \\
early-stopping          &  & 12                 & 12                & 25                & 14                & 11                \\
\midrule
\multicolumn{7}{@{}l}{U-Net}  \\
\midrule
2D                      &  & 0.741 (0.004)              & 0.479 (0.010)             & 0.898 (0.001)             & 0.537 (0.272)                & 0.788 (0.005) \\
proposed ($d=3$)        &  & 0.749 (0.005)              & 0.494 (0.015)             & 0.884 (0.006)             & 0.495 (0.137)                & 0.785 (0.005) \\
proposed ($d=5$)        &  & 0.743 (0.006)              & 0.486 (0.016)             & 0.892 (0.003)             & 0.423 (0.222)                & 0.847 (0.004) \\
proposed ($d=7$)        &  & 0.744 (0.003)              & 0.495 (0.009)             & 0.894 (0.004)             & 0.445 (0.208)                & 0.861 (0.004) \\
proposed ($d=9$)        &  & 0.736 (0.011)              & 0.505 (0.011)             & 0.896 (0.003)             & 0.404 (0.171)                & 0.840 (0.004) \\
proposed ($d=11$)       &  & 0.749 (0.005)              & 0.492 (0.011)             & 0.886 (0.008)             & 0.479 (0.081)                & 0.818 (0.004) \\
\midrule
channel-based ($d=3$)   &  & 0.736 (0.005)              & 0.489 (0.008)             & 0.903 (0.001)             & 0.667 (0.022)                & 0.789 (0.005) \\
channel-based ($d=5$)   &  & 0.742 (0.008)              & 0.487 (0.009)             & 0.903 (0.001)             & 0.512 (0.264)                & 0.855 (0.004) \\
channel-based ($d=7$)   &  & 0.742 (0.010)              & 0.477 (0.013)             & 0.901 (0.001)             & 0.504 (0.181)                & 0.859 (0.004) \\
channel-based ($d=9$)   &  & 0.739 (0.005)              & 0.481 (0.009)             & 0.903 (0.001)             & 0.519 (0.168)                & 0.853 (0.005) \\
channel-based ($d=11$)  &  & 0.742 (0.004)              & 0.485 (0.015)             & 0.902 (0.001)             & 0.501 (0.195)                & 0.792 (0.005) \\
3D                      &  & 0.724 (0.004)              & 0.436 (0.037)             & 0.699 (0.041)             & 0.318 (0.159)                & 0.827 (0.004) \\ 
\midrule
\multicolumn{7}{@{}l}{Seg-Net}  \\
\midrule
2D                      &  & 0.716 (0.003)              & 0.429 (0.010)             & 0.766 (0.002)             & 0.406 (0.145)                & 0.756 (0.005) \\
proposed ($d=3$)        &  & 0.719 (0.010)              & 0.435 (0.009)             & 0.778 (0.002)             & 0.395 (0.223)                & 0.816 (0.005) \\
proposed ($d=5$)        &  & 0.717 (0.007)              & 0.443 (0.009)             & 0.774 (0.002)             & 0.465 (0.229)                & 0.797 (0.005) \\
proposed ($d=7$)        &  & 0.718 (0.005)              & 0.443 (0.007)             & 0.772 (0.006)             & 0.409 (0.217)                & 0.809 (0.005) \\
proposed ($d=9$)        &  & 0.719 (0.009)              & 0.447 (0.010)             & 0.761 (0.008)             & 0.484 (0.242)                & 0.828 (0.005) \\
proposed ($d=11$)       &  & 0.721 (0.004)              & 0.444 (0.007)             & 0.767 (0.009)             & 0.449 (0.166)                & 0.751 (0.005) \\
\midrule
channel-based ($d=3$)   &  & 0.720 (0.004)              & 0.428 (0.012)             & 0.777 (0.002)             & 0.446 (0.136)                & 0.805 (0.005) \\
channel-based ($d=5$)   &  & 0.713 (0.005)              & 0.417 (0.015)             & 0.781 (0.002)             & 0.336 (0.193)                & 0.805 (0.005) \\
channel-based ($d=7$)   &  & 0.717 (0.004)              & 0.417 (0.008)             & 0.778 (0.003)             & 0.349 (0.251)                & 0.816 (0.005) \\
channel-based ($d=9$)   &  & 0.722 (0.006)              & 0.424 (0.013)             & 0.772 (0.004)             & 0.385 (0.147)                & 0.807 (0.005) \\
channel-based ($d=11$)  &  & 0.717 (0.002)              & 0.410 (0.016)             & 0.777 (0.003)             & 0.387 (0.239)                & 0.744 (0.005) \\
3D                      &  & 0.687 (0.016)              & 0.252 (0.091)             & 0.381 (0.054)             & 0.037 (0.017)                & 0.635 (0.005) \\
\bottomrule
\end{tabular}
\end{adjustbox}
\label{tab:performance_test_dice}
\end{table*}

\subsection{Supplementary Qualitative Results}
\begin{figure*}[!ht] 
    \centering
    \includegraphics[trim=5 0 10 0, clip, width=\textwidth]{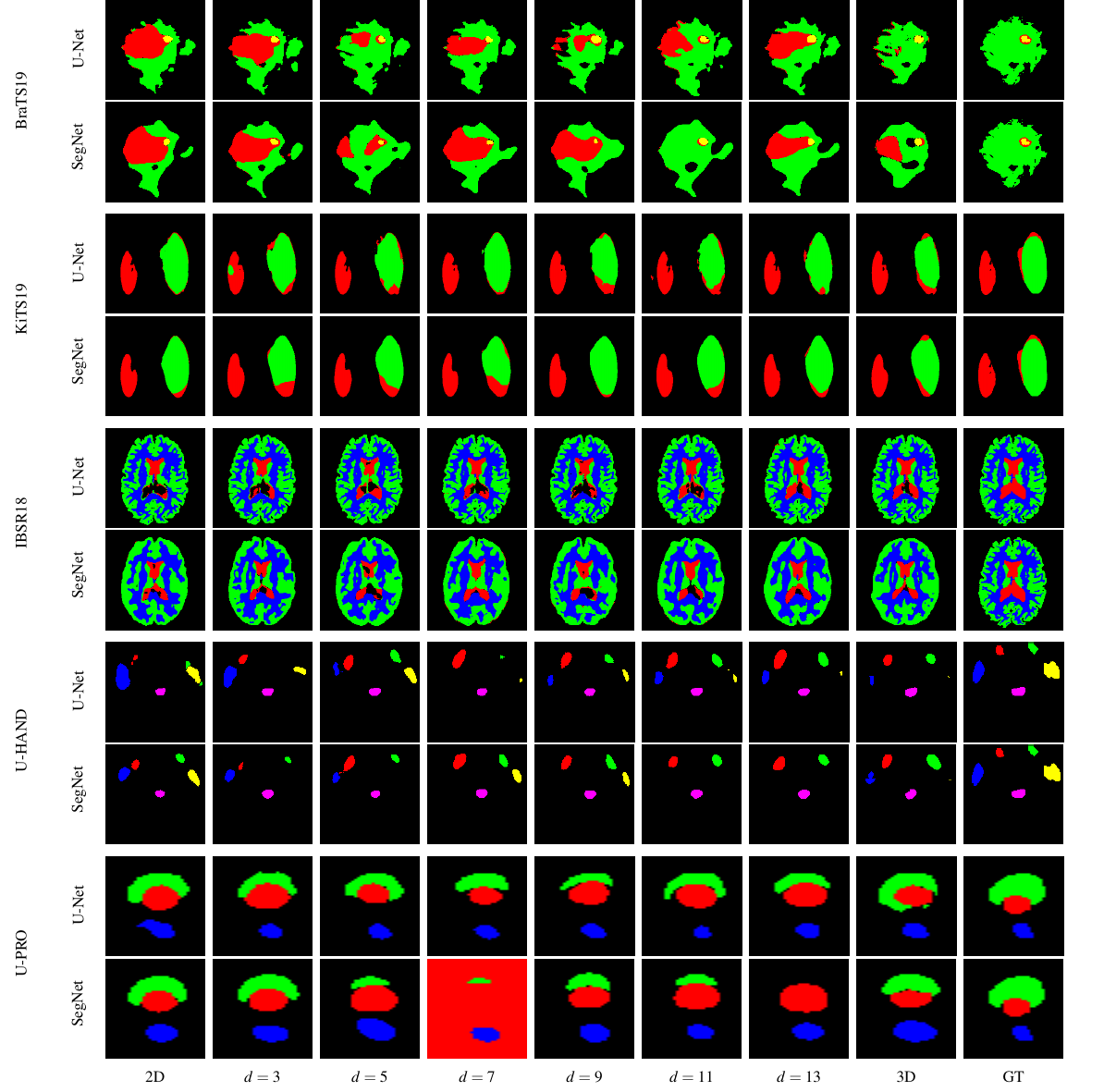}
    \caption{Qualitative results of the channel-based method on all data sets (see Figure 10 in the main paper for the qualitative results of the proposed method in the same evaluated examples). From top to bottom: (i) \acrshort{brats} tumor structures: the necrotic and non-enhancing tumor core (NCR/NET---label 1, red), the peritumoral edema (ED---label 2, green) and the GD-enhancing tumor (ET---label 4, yellow); (ii) \acrshort{kits} class structure: the kidney (red) and kidney tumor (green); (iii) \acrshort{ibsr} class structure: cerebrospinal fluid (red), white matter (green) and gray matter (blue); (iv) \acrshort{hene} class structure: left and right submandibular glands (red and green), left and right parotid glands (dark blue and yellow), larynx (light blue), and medulla oblongata (pink); (v) \acrshort{pros} class structure: prostate (red), bladder (green) and rectum (blue). From left to right: 2D, $d=3,5,7,9,11,13$, 3D, and ground truth (GT).}
    \label{fig:qualitative_channel_1}
\end{figure*}

\begin{figure*}[!ht] 
    \centering
    \includegraphics[trim=5 0 10 0, clip, width=\textwidth]{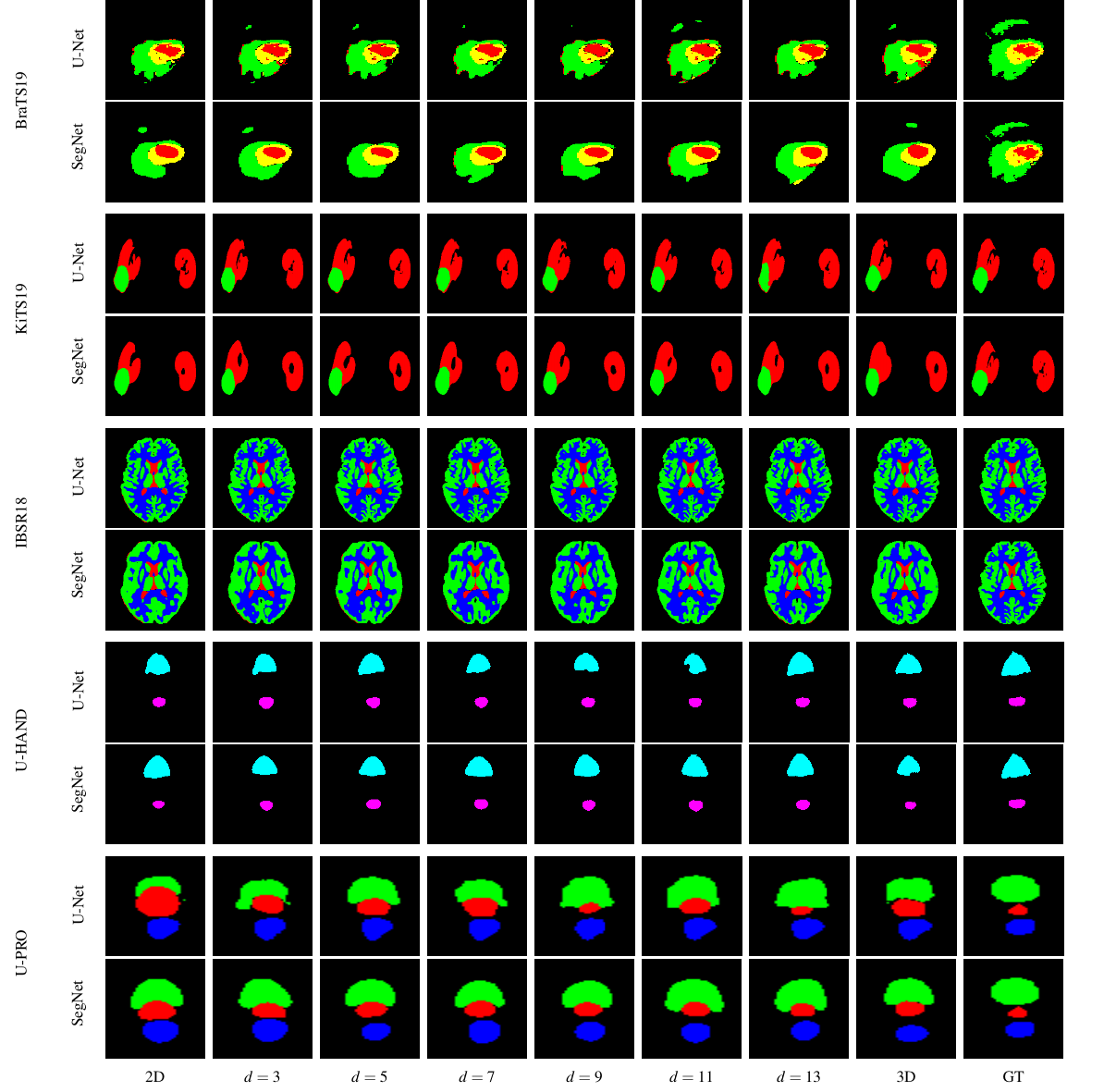}
    \caption{Qualitative results of proposed method on all data sets (see \figref{fig:qualitative_channel_2} in the Supplementary Material for the qualitative results of the channel-based method in the same evaluated examples). From top to bottom: (i) \acrshort{brats} tumor structures: the necrotic and non-enhancing tumor core (NCR/NET---label 1, red), the peritumoral edema (ED---label 2, green) and the GD-enhancing tumor (ET---label 4, yellow); (ii) \acrshort{kits} class structure: the kidney (red) and kidney tumor (green); (iii) \acrshort{ibsr} class structure: cerebrospinal fluid (red), white matter (green) and gray matter (blue); (iv) \acrshort{hene} class structure: left and right submandibular glands (red and green), left and right parotid glands (dark blue and yellow), larynx (light blue), and medulla oblongata (pink); (v) \acrshort{pros} class structure: prostate (red), bladder (green) and rectum (blue). From left to right: 2D, $d=3,5,7,9,11,13$, 3D, and ground truth (GT).}
    \label{fig:qualitative_proposed_2}
\end{figure*}

\begin{figure*}[!ht] 
    \centering
    \includegraphics[trim=5 0 10 0, clip, width=\textwidth]{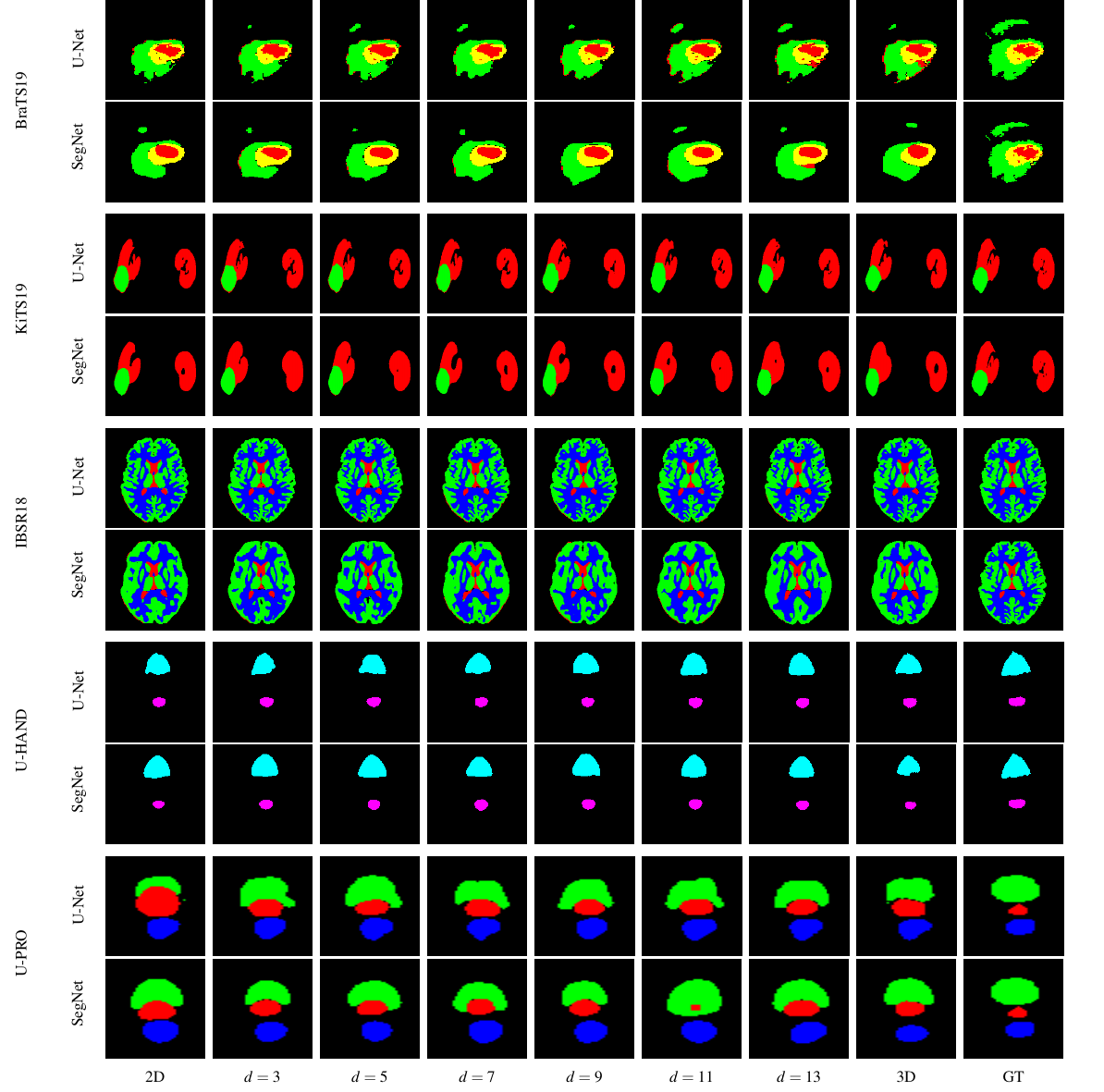}
    \caption{Qualitative results of channel-based method on all data sets (see \figref{fig:qualitative_proposed_2} in the Supplementary Material for the qualitative results of the proposed method in the same evaluated examples). From top to bottom: (i) \acrshort{brats} tumor structures: the necrotic and non-enhancing tumor core (NCR/NET---label 1, red), the peritumoral edema (ED---label 2, green) and the GD-enhancing tumor (ET---label 4, yellow); (ii) \acrshort{kits} class structure: the kidney (red) and kidney tumor (green); (iii) \acrshort{ibsr} class structure: cerebrospinal fluid (red), white matter (green) and gray matter (blue); (iv) \acrshort{hene} class structure: left and right submandibular glands (red and green), left and right parotid glands (dark blue and yellow), larynx (light blue), and medulla oblongata (pink); (v) \acrshort{pros} class structure: prostate (red), bladder (green) and rectum (blue). From left to right: 2D, $d=3,5,7,9,11,13$, 3D, and ground truth (GT).}
    \label{fig:qualitative_channel_2}
\end{figure*}

Example segmentations are illustrated in \figref{fig:qualitative_channel_1}--\ref{fig:qualitative_channel_2}.
It is important to emphasize that the images are randomly selected single slices from thousands of samples and are therefore presented purely for illustrative purposes, and might not always be a representation of the overall segmentation performance of a particular data set.

\end{document}